\RequirePackage{fix-cm}
\documentclass[twocolumn]{svjour3}          
\smartqed  
\usepackage{dblfloatfix}
\usepackage{float}
\usepackage{graphicx}
\usepackage{hyperref}
\usepackage[switch]{lineno}
\usepackage{wrapfig}
\usepackage{overpic}
\usepackage{times}
\usepackage{epsfig}
\usepackage{bm}
\usepackage{breakurl}
\usepackage{graphicx, subfigure}
\usepackage{amsmath,amssymb, amsfonts, euscript, mathrsfs}
\usepackage{subeqnarray}
\usepackage{cases}
\usepackage[sectionbib,round]{natbib}
\usepackage{algorithmic}

\usepackage[linesnumbered, ruled]{algorithm2e}
\SetKwRepeat{Do}{do}{while}%

\usepackage{fancybox}
\usepackage{picins}
\usepackage{picinpar} 
\usepackage[table]{xcolor}
\usepackage{wrapfig}
\definecolor{lightgray}{gray}{0.98}
\definecolor{lightblue}{rgb}{0.98,0.98,1.0}
\usepackage{subfigure}
\usepackage{multirow}
\usepackage{booktabs}
\usepackage{ulem}
\usepackage{comment}
\usepackage{listings}
\allowdisplaybreaks[4]

\definecolor{mygreen}{rgb}{0,0.6,0}
\definecolor{mygray}{rgb}{0.5,0.5,0.5}
\definecolor{mymauve}{rgb}{0.58,0,0.82}
\newcommand{\executeiffilenewer}[3]{%
\ifnum\pdfstrcmp{\pdffilemoddate{#1}}%
{\pdffilemoddate{#2}}>0%
{\immediate\write18{#3}}\fi%
}

\makeatletter
\DeclareRobustCommand\bigop[1]{%
  \mathop{\vphantom{\sum}\mathpalette\bigop@{#1}}\slimits@
}
\newcommand{\bigop@}[2]{%
  \vcenter{%
    \sbox\z@{$#1\sum$}%
    \hbox{\resizebox{\ifx#1\displaystyle.9\fi\dimexpr\ht\z@+\dp\z@}{!}{$\m@th#2$}}%
  }%
}
\makeatother

\newcommand{\stkout}[1]{\ifmmode\text{\sout{\ensuremath{#1}}}\else\sout{#1}\fi}

\allowdisplaybreaks[4]

\modulolinenumbers[1]
\journalname{}
\begin{document}\sloppy

\title{OpenTM: An Open-source, Single-GPU, Large-scale Thermal Microstructure Design Framework}

\subtitle{}


\author{Yuchen Quan\ $^{1}$  and
        Xiaoya Zhai$^{2\star}$ \and
        Xiao-Ming Fu$^{2}$ 
}


\institute{
        \at
        $^1$School of Artificial Intelligence and Data Science, University of Science and Technology of China, Hefei, China \\
        $^2$School of Mathematical Sciences, University of Science and Technology of China, Hefei, China \\
        \at
        $^\star$Corresponding Author. \email{xiaoyazhai@ustc.edu.cn}
}

\date{Received: date / Accepted: date}

\maketitle

\begin{abstract}
Thermal microstructures are artificially engineered materials designed to manipulate and control heat flow in unconventional ways. 
This paper presents an educational framework, called \emph{OpenTM}, to use a single GPU for designing periodic 3D high-resolution thermal microstructures to match the predefined thermal conductivity matrices with volume fraction constraints. 
Specifically, we use adaptive volume fraction to make the Optimality Criteria (OC) method run stably to obtain the thermal microstructures without a large memory overhead.
Practical examples with a high resolution $128 \times 128 \times 128$ run under 90 seconds per structure on an NVIDIA GeForce GTX 4070Ti GPU with a peak GPU memory of 355 MB. 
%
Our open-source, high-performance implementation is publicly accessible at \url{https://github.com/quanyuchen2000/OPENTM}, and it is easy to install using Anaconda.
Moreover, we provide a Python interface to make OpenTM well-suited for novices in C/C++.
\end{abstract}

\section{Introduction}\label{sec:intro}

\begin{figure*}
	\centering
	\begin{overpic}[width=0.99\linewidth]{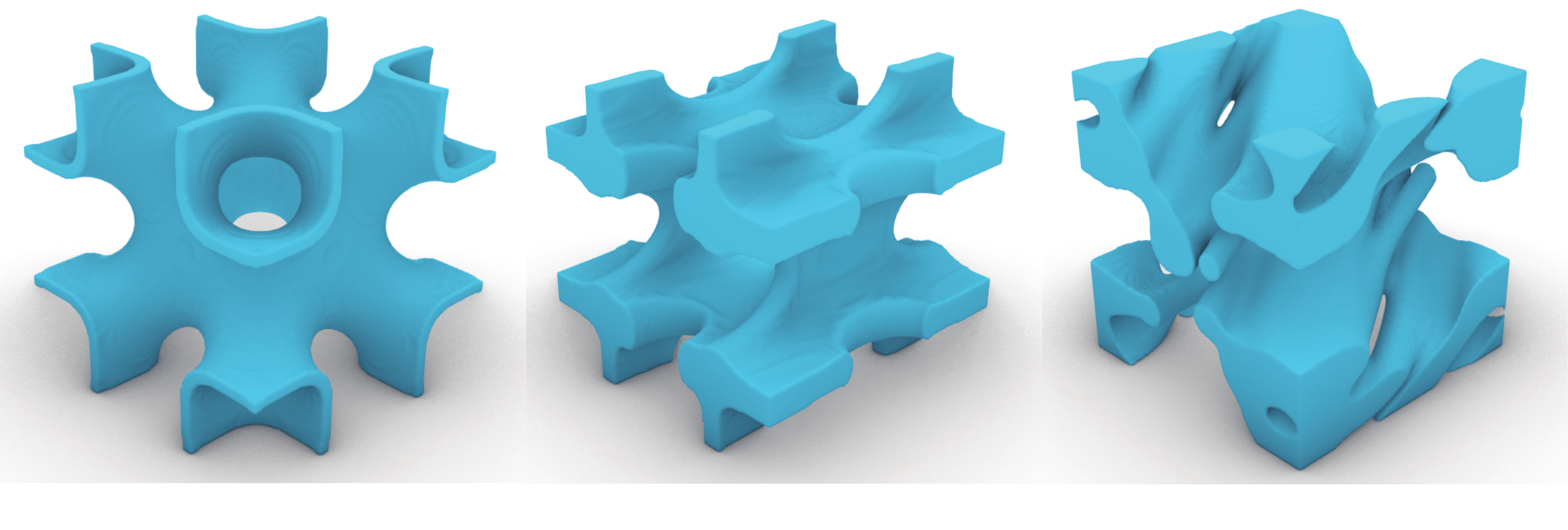}
		{
			\put(12,1){ (a) Isotropy }
			\put(40,1){ (b) Orthotropic anisotropy }
			\put(78,1){ (c) Anisotropy }
		}
	\end{overpic}
	\vspace{-2mm}
	\caption{
		Designing a large-scale thermal microstructure on a unit cell domain ( $128\times128\times128$ elements) with three different thermal conductivity tensor types:(1) isotropy, (2) orthotropic anisotropy, and (3) anisotropy. The target tensors from left to right are $[0.1, 0.1, 0.1, 0, 0, 0]$, $[0.3, 0.2, 0.1, 0, 0, 0]$, $[0.3, 0.2, 0.1, 0.1, 0.05, 0.05]$. We represent tensors using vectors as~\eqref{eq:rep}.
	}
	\label{fig:teaser}
\end{figure*}

Thermology represents one of the fundamental natural phenomena. Heat is predominantly transferred through conduction~\cite{chung2001materials}, convection~\cite{lappa2009thermal}, and radiation~\cite{howell2020thermal}, with heat flux serving as a manifestation of heat conduction. 
A non-uniform spatial thermal conductivity distribution enables artificial control of heat flow direction on a macroscopic scale. 
In recent years, deliberate manipulation of heat flux has garnered significant attention~\cite{li2021transforming}. Given the theories~\cite{tritt2005thermal} capable of calculating the thermal conductivity matrix for each spatial point and facilitating directional guidance of heat flux, designing anisotropic thermal materials merits further exploration.

Thermal metamaterial design entails an inverse design approach, where the designer predefines the desired extraordinary thermal functionality. 
The judicious distribution of materials then realizes artificial control over heat conduction. 
Topology optimization is a powerful and effective method for finding microstructures with uncommon properties, such as negative Poisson's ratio and high bulk modulus with low volume fraction.
Currently, researchers integrate topological optimization methods to design a range of thermal microstructures, including structures for heat flow shielding and focusing inversion~\cite{ji2021designing,dede2014thermal}. 
These designs are also tailored to address multi-physics or multifunctional challenges~\cite{fujii2019optimizing,fujii2020cloaking}. 
Despite the extensive research on thermal structures, there remains a notable scarcity of open-source frameworks dedicated to thermal metamaterial design.


We focus on designing high-resolution periodic 3D thermal microstructures via the density-based method. 
Specifically, we aim to design a microstructure with a specified thermal conductivity matrix to manipulate thermal flux. 
The theory of thermal metamaterial design through topology optimization is well-established and has been used to fabricate thermal cloaking devices~\cite{sha2021robustly}; however, the prevailing focus in current research remains largely on 2D thermal microstructures and cannot meet the demand for 3D microstructure in practical applications.
To bridge this gap, the educational aim of this paper is to provide a user-friendly open-source framework with a Python interface for designing 2D and 3D high-resolution thermal metamaterials.


The rapid advancement of topology optimization owes much to the contributions of open-source code initiatives. 
The inception of topology optimization can be traced back to the homogenization method proposed by~\cite{bendsoe1988generating}. 
Subsequently, a suite of classic topology optimization techniques, such as the Solid Isotropic Material with Penalization (SIMP) method~\cite{bendsoe1999material,stolpe2001alternative,bendsoe1989optimal}, Evolutionary Structural Optimization (ESO)~\cite{querin1998evolutionary}, the level set method~\cite{wang2003level,van2013level}, and Moving Morphable Components (MMC) methods~\cite{zhang2016new}, have been developed. 
Importantly, these methods are accompanied by open-source frameworks and educational papers, 
such as Top99~\cite{sigmund200199} and Top88 Matlab code~\cite{andreassen2011efficient} for SIMP method, \cite{picelli2021101} for the BESO method, 
designed to facilitate a deeper understanding for newcomers. 
\cite{wang2023open} also summarize the open source codes to help beginners start their topology optimization journeys.

The design space directly impacts the performance of final topology optimization results. 
In our focused density-based structures, resolution is critical. 
Expanding the design space improves the results but significantly increases the computational load.
As a result, researchers have devised a range of topology optimization algorithms tailored for high-performance computing.
\cite{aage2015topology} introduce PETSc, a high-resolution multi-CPU framework for topology optimization. 
\cite{7332965} propose a multigrid solver to solve high-resolution topology optimization problems. 
\cite{SOTIROPOULOS2020441} present the topology optimization module of a high-performance optimization computing platform targeting civil engineering problems. 
\cite{TRAFF2021107349} introduce an efficient parallel geometric/algebraic multigrid precondition for high contrast shell problems dealing with more than 11 million linear shell elements. 
\cite{HERREROPEREZ2021103006} propose a multi-GPU framework with mixed-precision techniques to solve large-scale topology optimization problems. 

In the metamaterial design, \cite{andreassen2014determine} introduce the numerical homogenization method for calculating the properties of composite materials. 
Subsequently, \cite{xia2015design} release an open-source framework for designing 2D mechanical microstructures using an energy homogenization approach. 
PETSc can also be used to design high-resolution 3D mechanical microstructures with multiple CPUs.
LIVE3D framework developed by \cite{zhang2023optimized} is an open-source framework for designing high-resolution 3D mechanical microstructures using a single GPU. This advancement eliminates the reliance on high-performance computing centers, making it feasible to design high-resolution microstructures on standard desktop computers. However, most of these frameworks are predominantly focused on mechanical metamaterial designs. 
Despite the refinement of the theory surrounding thermal microstructures~\cite{yang2021controlling}, there remains a notable absence of open-source frameworks tailored for their inverse design. 
Thus, our educational goal is to fill this gap.

In this paper, we present an educational framework, called \emph{OpenTM}, for the optimal design of materials to match a given thermal conductivity tensor. 
To make the Optimality Criteria (OC) method applicable to thermal microstructure design, we propose changing the upper volume fraction bound adaptively.
%
Our implementation extends the LIVE3D framework~\cite{zhang2023optimized} and uses various acceleration algorithms within that framework such as multi-grid acceleration.
Consequently, we can optimize the high-resolution thermal microstructure on standard computers with only a single GPU.
Practical high-resolution examples with about 2.1 million ﬁnite elements run less than 0.18 seconds per iteration on an NVIDIA GeForce GTX 4070Ti GPU with a peak GPU memory of 355 MB (Fig.~\ref{fig:teaser}).

The main advantages of OpenTM are as follows.
(1) The OC method is made available for thermal microstructure design, thereby not consuming too much memory.
(2) It can run with a single GPU, significantly increasing its application range. 
(3) Our framework is easy to use in three aspects. It is easy to install using Anaconda, easy to use with a Python interface, and easy to customize the thermal microstructure with automatic differentiation.
%
%
This capability facilitates a more comprehensive exploration of thermal metamaterials for users. 
The proposed framework, including a detailed installation video, is publicly available for download at \url{https://github.com/quanyuchen2000/OPENTM}.
%


\section{Thermal microstructure optimization}\label{sec:method}
\subsection{General thermal conduction equation}
The governing partial differential thermal conduction equation of the steady-state case for the temperature field $\bm{T}$ is 
\begin{align}
	&\nabla \cdot (k\nabla \bm{T}) + \bm{f} = 0 \ \,  \text{in} \, \  \Omega \label{eq:gen_kt}\\
	&\bm{T} = 0 \ \, \text{on} \, \  \Gamma_D \label{eq:tb}\\
	&(k\nabla \bm{T})\cdot \bm{n} = 0 \ \, \text{on} \, \  \Gamma_N \label{eq:tn}
\end{align}
where $\Omega \subset \mathbb{R}^3$ is the spatial design domain with the boundary $\Gamma = \Gamma_D \cup \Gamma_N$, $\Gamma_D $ and $ \Gamma_N$ are Dirichlet boundary and Neumann boundary, $k$ is the heat conduction coefficient, $\bm{f}$ is the volumetric thermal load, and $\bm{n}$ is an outward unit normal vector. The heat flux is $\mathbf{q} = k \nabla \bm{T}$. 
Through the deduction~\cite{gersborg2006topology}, we can get the finite element equilibrium equation:
\begin{equation}
	\bm{K}\bm{T} = \bm{f} \label{eq:ktf}
\end{equation}
where $\bm{K}$ is the symmetric thermal conductivity matrix. 

\subsection{Homogenization}

Homogenization theory~\cite{} aims to determine the macroscopic equivalent properties of a microstructure based on its material distribution. 
Analogously, we can utilize the material density distribution $\bm{\rho}$ in thermal microstructures to calculate the equivalent heat conduction tensor $\bm{\kappa}^H$. 
Periodic boundary conditions are introduced to ensure an infinite array. 
\begin{equation}
	\begin{aligned}
		& \bm{T}(\bm{x}+\bm{w}) = \bm{T} (\bm{x}) + \bm{G}\bm{w} \quad \bm{x} \in \partial \Omega_v, \\
		& \bm{q}(\bm{x}+\bm{w}) = -\bm{q} (\bm{x}) \quad \bm{x} \in \partial \Omega_v \label{eq:kp}.
	\end{aligned}
\end{equation}
where the temperature $\bm{T}$ is periodic function whose period is $\bm{w}$ and $\bm{G}$ is the macro temperature gradient and $\bm{q}$ represents the directed heat flux passing through the boundary. 
The homogenized thermal conductivity matrix $\kappa_{ij}^{H}$ is determined as:
\begin{equation}\label{eq:homo}
	\kappa_{ij}^{H} = \frac{1}{|\Omega_v|}\int_{\Omega_v}(q_r^{0(i)} - q_r^{(i)})\kappa_{rs}(q_s^{0(j)} - q_s^{(j)})d\Omega_v.
\end{equation}
Here $|\Omega_v|$ is the volume of the microstructure, $i, j = \{1,2,3\}$ represent indices in three dimensions, and $q^{0(i)}_r$ is the test heat flux given at first. 
In the 3D case, there are 3 identical test heat fluxes along 3 coordinate axes. 
$q^{(i)}_r$ is the unknown heat flux inside the structure after applying $q^{0(i)}_r$ and can be solved by the equilibrium equation: 
\begin{equation}
	\int_{\Omega_v} q_r^{(i)}\kappa_{rs} q_s(v)d\Omega_v = \int_{\Omega_v} q_r^{0(i)}\kappa_{rs}q_s(v)d\Omega_v,
\end{equation}
where $v$ represents the virtual temperature field.

The Finite Element Method is used for numerical computation to solve~\eqref{eq:ktf} and~\eqref{eq:kp}. 
The microstructure design domain $\Omega_v$ is discretized into $M$ elements. 
So, the thermal conductivity matrix is computed as
\begin{align}
	&\bm{K}\bm{T}^{(i)} = \bm{f}^{(i)} \label{eq:kti},
\end{align}
where $\bm{f}^{(i)}$ the test thermal load alone the 3 axis. 
The numerical homogenized thermal conductivity matrix $ \kappa_{ij}^H$ is:
\begin{equation}
	\kappa_{ij}^H = \frac{1}{|\Omega_v|}\sum\limits_e (T_e^{0(i)}-T_e^{(i)}) \kappa_e(\rho_e) (T_e^{0(j)}-T_e^{(j)}),
\end{equation}
where $T_e^0$ is the nodal temperature defined on element $e$ which is calculated by the test flux, $T_e$ is from the result of~\eqref{eq:kti} and $\kappa_e(\rho_e)$ is based on the SIMP approach~\cite{bendsoe1989optimal}, $\kappa_e(\rho_e) = \kappa_{\min} + \rho_e^p (\kappa^0-\kappa_{\min})$. $\kappa^0$ is the thermal conductivity coefficient of the solid material. $\kappa_{\min}$ is usually taken as $1e^{-5}$.

\begin{figure*}[t]
	\centering
	\begin{overpic}[width=0.99\linewidth]{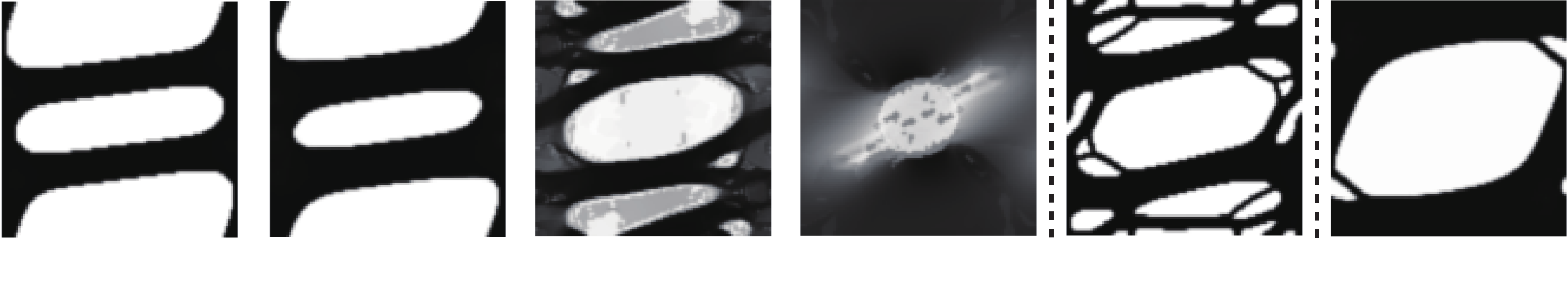}
		{
			\put(0,1){$g:$}
			\put(5,1){ $0.0138$ }
			\put(22,1){ $2.01e^{-4}$ }
			\put(39,1){ $5.46e^{-7}$ }
			\put(56,1){ $2.68e^{-9}$ }
			\put(73,1){ $9.98e^{-7}$ }
			\put(90,1){ $1.99e^{-5}$ }
			\put(6,-0.5){ \textbf{(a) }}
			\put(23,-0.5){ \textbf{(b) }}
			\put(40,-0.5){\textbf{(c) }}
			\put(57,-0.5){ \textbf{(d) }}
			\put(74,-0.5){ \textbf{(e) }}
			\put(91,-0.5){\textbf{ (f)} }
		}
	\end{overpic}
	\vspace{0mm}
	\caption{
		\textbf{(a) $\sim$ (d)} shows the results obtained by solving the model~\eqref{md:model1} with the volume fraction of 40\% 50\% 60\% 70\%. 
		The numerical value beneath each subfigure corresponds to the objective function $g$. 
		\textbf{(e)} and \textbf{(f)} shows the results of solving the models~\eqref{md:model2} and~\eqref{md:model3} with volume fraction of 53.3\% and 50.6\%.
	}
	\label{fig:model1}
\end{figure*}
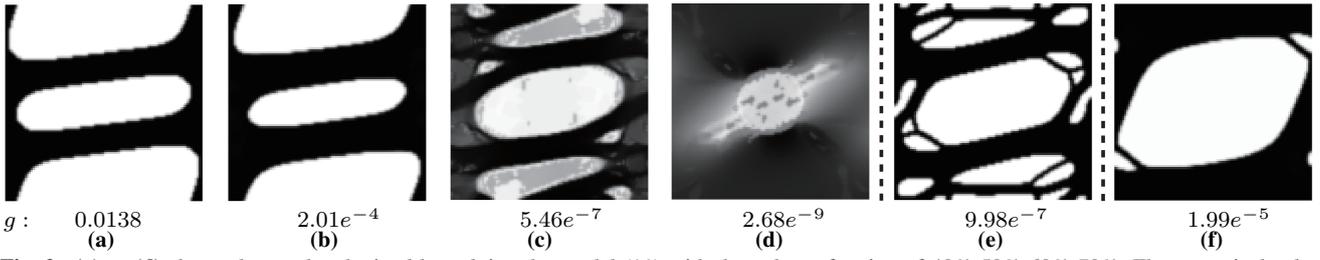

Since $\bm{\kappa}$ is a 3x3 symmetric matrix, when we refer to the target thermal conductivity tensor as a 1x6 vector later in the paper, the assumed correspondence is:
\begin{equation}
	\kappa_{3\times3} \rightarrow [\kappa_{11}^H, \kappa_{22}^H, \kappa_{33}^H, \kappa_{12}^H, \kappa_{23}^H, \kappa_{13}^H].
	\label{eq:rep}
\end{equation}

\subsection{Optimization models}
The inverse homogenization problem for thermal microstructure design is performed on a unit domain $\Omega = [0,1]^3$, which is evenly discretized into $M$ elements. 
Each element is assigned a density variable $\rho_e$ and a fixed volume $v_e$. 
All density variables $\rho_e\,(e = 1,\cdots, M)$ form a vector $\bm{\rho}$. 
To design a structure with a certain thermal conductivity matrix $\bm{\kappa}^*$, the optimization of thermal microstructure can be formulated as follows:
\begin{equation}
	\begin{aligned}
		\min_{\bm{\rho}} \quad & g(\bm{\kappa}^H(\bm{\rho}), \bm{\kappa}^*), \\
		s.t. \quad & \bm{K(\rho)T} = \bm{f}, \\
		& V(\bm{\rho}) = \frac{\sum\limits_{e} v_e \cdot \Tilde{\rho}_e}{|\Omega|} \leq V,\\
		& 0 \leq \rho_e \leq 1.
	\end{aligned}
	\label{md:model1}
\end{equation}
Here, the function \( g(\cdot, \cdot) \) is distance function where \( g(a, b) = ||a(:) - b(:)||^2, \, a(:)\) converts a tensor $a$ to a column vector. 
Other functions are discussed in Section~\ref{sub:ext}. 
$V$ is a given volume fraction threshold. 
To avoid checkboard patterns, the density $\rho_e$ is filtered in its neighboring regions $N_e$ to obtain the smoothed density $\Tilde{\rho}_e $:
\begin{equation}
	\Tilde{\rho}_e = \frac{\sum_{i\in N_e} d_e(x_i,r)v_i\rho_i}{\sum_{i\in N_e}d_e(x_i,r)v_i},
\end{equation}
where $v_i$ is the element's volume, $r$ is a predefined filter radius, $d_e(x_i,r)$ is a weight function.

\begin{figure}[t]
	\centering
	\begin{overpic}[width=0.99\linewidth]{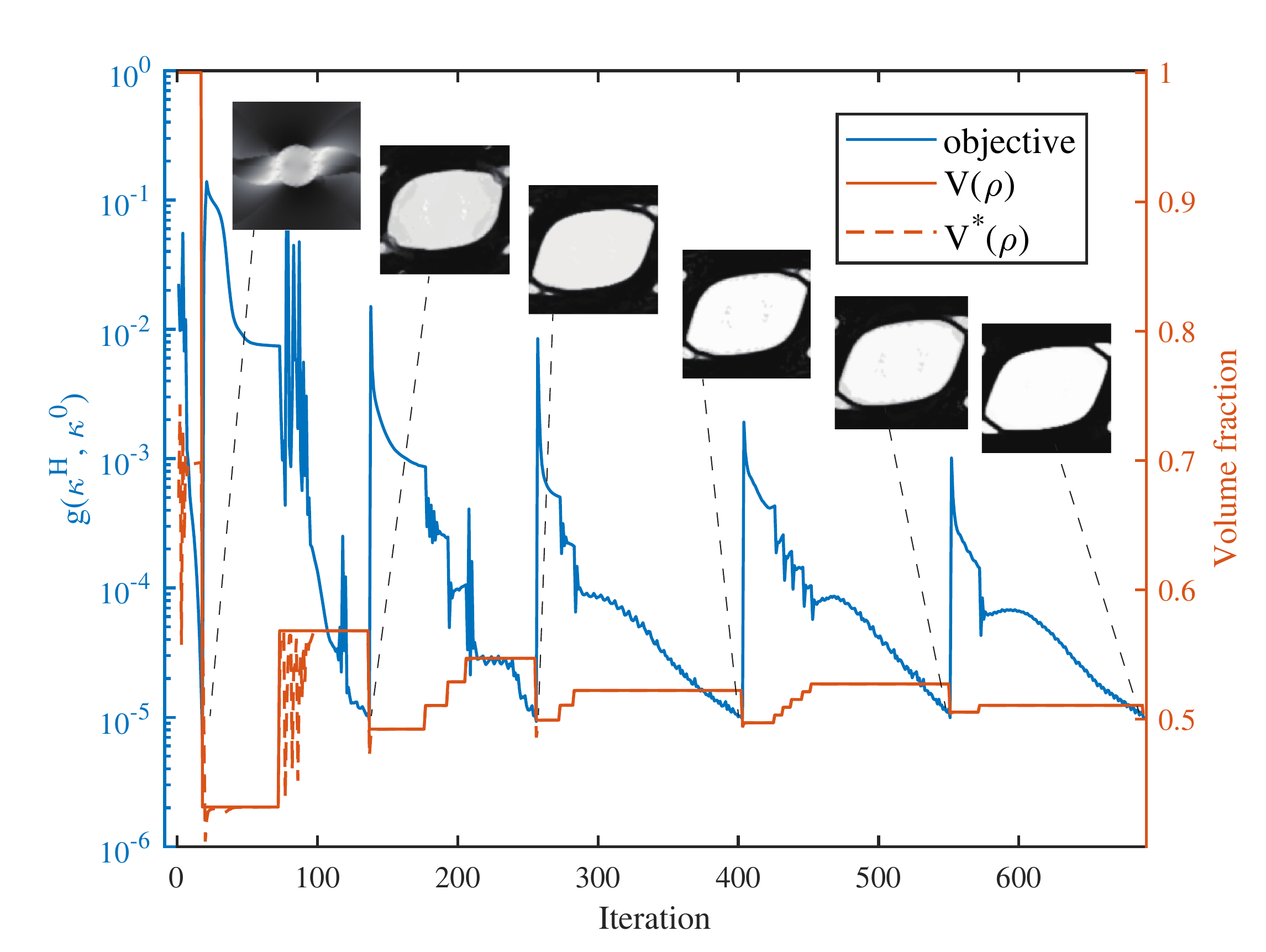}
		{
		}
	\end{overpic}
	\vspace{0mm}
	\caption{Iteration curve of Fig.~\ref{fig:model1}~(f) by solving the model~\eqref{md:model3}. 
	}
	\label{fig:loss func}
\end{figure}

Nevertheless, this formulation demonstrates limitations in its performance, occasionally resulting in the failure to produce black-white microstructures. 
We take a 2D region with $100\times100$ elements as an example. The predefined $\bm{\kappa}^* = \left[ \begin{array}{ccc}  
	\kappa^*_{11} & \kappa^*_{12} \\ 
	\kappa^*_{21} & \kappa^*_{22} \\  
\end{array} \right]=
\left[ \begin{array}{ccc}  
	0.4 & 0.05 \\ 
	0.05 & 0.2 \\  
\end{array} \right]
$ and  \(\kappa^0 = 1\), a density filter ratio $r = 2$. When testing different volume fractions shown in Fig.~\ref{fig:model1}~(a)-(d), 
the volume fraction of these four results all reach the upper limits. 
As the volume fraction increases, the smaller the objective function $g$, while 
the optimized results end up with gray areas even when the projection operation is applied. 
In contrast to mechanical microstructure optimization, the volume fraction of thermal microstructure exhibits no direct correlation with the objective function. 
Specifically, the thermal microstructure permits the volume constraint to remain below its upper limit even after attaining the objective function. 
Thus, density is under-utilized, resulting in the generation of gray areas.

From this point of view, another optimization problem, which is also used in \cite{sha2022topology}, is presented by modifying the objective function and constraints:
\begin{align}
	\min_{\bm{\rho}} \quad & V(\bm{\rho}) = \frac{\sum\limits_{e} v_e \cdot \Tilde{\rho}_e}{|\Omega_v|}, \notag\\
	s.t. \quad & \bm{K(\rho)T} = \bm{f}, \label{md:model2}\\
	&  g(\bm{\kappa}^H(\bm{\rho}), \bm{\kappa}^*) \leq \epsilon, \notag\\
	& 0 \leq \rho_e \leq 1. \notag
\end{align}
Fig.~\ref{fig:model1}~(e) shows the results generated by solving~\eqref{md:model2}, which tends to employ black-white density distributions by minimizing volume fractions while the pre-specified parameters are all consistent with~\eqref{md:model1}. 
This approach has proved effective in most 2D cases and small-scale 3D scenarios. 
However, for large-scale problems (exceeding \(128^3\) voxels), the Method of Moving Asymptotes (MMA)~\cite{svanberg1987method} necessitates significant memory allocation for intermediate parameters during auxiliary computations, rendering it impractical due to excessive memory consumption. 
Furthermore, the problem~\eqref{md:model2} cannot be solved by the Optimality Criteria (OC) method. 
OC method requires the evaluation of the constraint function to update the Lagrange multipliers, making the computation of \(g(\bm{\kappa}^H(\bm{\rho}), \bm{\kappa}^*)\) computationally intensive. 
%
Therefore, we propose a new optimization problem as an alternative:
\begin{align}
	\begin{aligned}
		\min_\rho \quad & g(\bm{\kappa}^H(\bm{\rho}), \bm{\kappa}^*), \\
		s.t. \quad & \bm{K(\rho)T} = \bm{f}, \\
		& V(\bm{\rho}) = \frac{\sum\limits_{e} v_e \cdot \Tilde{\rho}_e}{|\Omega_v|} \leq V^*(\bm{\rho}),\\
		& 0 \leq \rho_e \leq 1.
	\end{aligned}
	\label{md:model3}
\end{align}
To enhance optimization efficiency while mitigating the occurrence of gray areas in the final output, we dynamically adjust the volume fraction threshold $V^*(\bm{\rho})$. 
Intuitively, when the objective function value is sufficiently low, we compute the number of partially filled voxels. 
A high number indicates that the volume fraction needs reduction. 
Conversely, if the objective function fails to decrease enough due to the volume constraint, we should consider increasing the volume fraction limit. 
We update $V^*(\bm{\rho})$ with these guidelines.
Algorithm~\ref{alg:uvf} shows the rules for updating the volume threshold $V^*$ (see more details in the provided open-source code). 
Through continuous adaptation of the volume fraction threshold, the density can be utilized more efficiently, thereby mitigating the emergence of gray areas resulting from density redundancy. 
Fig.~\ref{fig:model1} (f) shows the final results, and Fig.~\ref{fig:loss func} shows the iteration curves. 
The algorithm terminates if the change in the objective function is smaller than the specified threshold  for three consecutive iterations. 

\IncMargin{0.5em}
\begin{algorithm}[t]
	\caption{Update $V^*(\bm{\rho})$ in  model~\eqref{md:model3}}\label{alg:uvf}
	\SetCommentSty{mycommfont}
	\SetKwInOut{AlgoInput}{Input}
	\SetKwInOut{AlgoOutput}{Output}
	\SetKwFunction{Init}{InitializePopulation}
	\SetKwFunction{enf}{enforceMacroLoad}
	\SetKwFunction{fem}{FEM}
	\SetKwFunction{sens}{SensitivityAnalysis}
	\SetKwFunction{updateRho}{updateDensity}
	\AlgoInput{density variable $\boldsymbol{\rho}$}
	\AlgoOutput{new Volume constraint $V^*$}
	Initialization: Decrease factor: $Df = 1$;  $V^* = 1$; Iteration $I$ = 1; $count = 0$; $I_{\max} = 500$; $g^{[0]}=1$; \\
	\For{$I \leq I_{\max}$}
	{
		$InvHomo()$;\tcp*[f]{Topology optimization}\\
		\If{$g^{[I]}\leq bound$}
		{
			$lowBound\gets mean(\boldsymbol{\rho}^{p})$; \\
			$gap \gets V^* - lowBound$;\\ 
			$V^* = V^* - gap * Df$;\tcp*[f]{Reduce $V^*$}\\
			$Df = 0.8*Df$;
		}
		$cond1 \gets is\_small(g^{[I-1]} - g^{[I]})$;\tcp*[f]{Little progress}\\
		$cond2 \gets is\_big(g^{[I]})$;\tcp*[f]{Need reduction}\\
		$cond3 \gets is\_small(V - V^*)$;\tcp*[f]{Reach boundary}\\
		\If{$cond1\, \mathbf{and}\, cond2\, \mathbf{and}\, cond3$} {
			$count = count +1$;
		}
		\Else{
			$count = 0$;
		}
		\If{$count \geq 5$} {
			$V^* = V^* + 0.3*gap* Df$;\tcp*[f]{Rebound $V^*$}\\
			$count = 0$;
		}
		$I = I+1$;
	}
\end{algorithm}
\DecMargin{0.5em}

For sensitivity analysis, the gradient with respect to element density $\rho_e$ is computed as
\begin{equation}
	\frac{\partial g}{\partial \rho_e} = \sum\limits_{ij} \frac{\partial g}{\partial \kappa_{ij}^H} \cdot \frac{\partial \kappa_{ij}^H}{\partial \Tilde{\rho}_e}\cdot \frac{ \Tilde{\rho}_e}{\partial \rho_e},
\end{equation}
where $\frac{\partial \kappa_{ij}^H}{\partial \Tilde{\rho}_e}$ is calculated as
\begin{equation}\label{eq:sens}
	\frac{\partial \kappa_{ij}^H}{\partial \Tilde{\rho}_e} = \frac{1}{|\Omega_v|}\sum\limits_e p\Tilde{\rho}_e^{p-1}(T_e^{0(i)}-T_e^{(i)}) \kappa^0 (T_e^{0(j)}-T_e^{(j)})
\end{equation}
Since the entire sensitivity analysis process follows the chain rule, we can use automatic differentiation to help us implement it.

\IncMargin{0.5em}
\begin{algorithm}[t]
	\caption{Topology optimization for thermal materials}\label{alg:ihm}
	\SetCommentSty{mycommfont}
	\SetKwInOut{AlgoInput}{Input}
	\SetKwInOut{AlgoOutput}{Output}
	\SetKwFunction{Init}{InitializePopulation}
	\SetKwFunction{enf}{enforceMacroLoad}
	\SetKwFunction{fem}{FEM}
	\SetKwFunction{sens}{SensitivityAnalysis}
	\SetKwFunction{updateRho}{updateDensity}
	\AlgoInput{ initial density vector $\boldsymbol{\rho}$}
	\AlgoOutput{optimized density vector $\boldsymbol{\rho}^*$}
	\For{each iteration}
	{
		$\mathbf{f}\gets\enf(\boldsymbol{\rho})$;\\
		$\mathbf{T}\gets\fem(\boldsymbol{\rho},\mathbf{f})$ by solving ;\\
		$\frac{\partial f}{\partial\rho_e}\gets\sens(\mathbf{u},\boldsymbol{\rho})$;\\
		$\boldsymbol{\rho}\gets\updateRho(\boldsymbol{\rho},\frac{\partial f}{\partial \rho_e})$;
	}
	$\boldsymbol{\rho}^*\leftarrow \boldsymbol{\rho}$;
\end{algorithm}
\DecMargin{0.5em}

\subsection{Solver}
Though the construction of a multi-grid method framework on GPU is inherited from \cite{zhang2023optimized}, a brief summary of the implementation is given for the completeness of the paper. The outline of the solver is shown in Algorithm \ref{alg:ihm}. 
\begin{itemize}
	\item \lstinline{enforceMacroLoad} The response thermal load on the vertex $v$ from a unit linear temperature gradient is:
	\begin{equation}
		f_v^i = \sum_{e=0}^{7} \rho_v^p f_{[e,i]}^0
	\end{equation}
	where $f_v^i$ is the thermal load on $v$ of the i-th direction temperature gradient. $f^0$ is the $8\times3$ equivalent load matrix calculated for the standard unit voxel. Since $f^0$ is pre-calculated and stored as a template, the calculation of the equivalent thermal loads for each node is independent and can be efficiently parallelized.
	\item \lstinline{FEM}
	To achieve better GPU parallelism in solving \eqref{eq:ktf}, we introduced multigrid solver constructed as Algorithm\ref{alg:multigrid} and within one relaxation we apply an 8-color Gauss-Seidel iteration.Central to the procedures is to compute $[KT]$ of any vertex $v$:
	\begin{equation}
		[KT]_v = \sum_{e=0}^7(\sum_{v_j=0}^7 K_{[7-e,v_j]}^0 T^{v_j,e})
	\end{equation}
	where $e$ is the incident element of $v$, $K^0$ is the element assembly matrix we obtained, $T^{v_j,e}$ means the temperature of $v_j$ vertex of element $e$. The residual is:
	\begin{equation}
		r_v = f_v - [KT]_v
	\end{equation}
	We introduce two notations for the relaxation:
	\begin{equation}
		M_v = \sum_{e=0}^{7} \rho^e \left( \sum_{v_j=0, v_j \neq 7-e}^{7} K^0_{[7-e, v_j]} T^{v_j,e} \right),
	\end{equation}
	
	\begin{equation}
		S_v = \sum_{e=0}^{7} \rho^e K^0_{[7-e, 7-e]},
	\end{equation}
	where we use \( M_v \) to denote the modified \([KT]_v\) and \( S_v \)
	to denote the sum of \(K^0\). Then, the Gauss-Seidel relaxation is performed via the following linear system to update \(T_v\):
	\begin{equation}
		S_vT_v = f_v - M_v
	\end{equation}
	Due to the 8-color Gauss-Seidel relaxation, all $T^{v_j,e}$ are constant when updating $T_v$. Therefore, the entire FEM process can be efficiently parallelized on the GPU, avoiding the significant memory overhead associated with assembling the global stiffness matrix.
	\item \lstinline{SensitivityAnalysis }
	Based on \eqref{eq:sens} and the FEM result above, it is evident that the calculation of element-wise sensitivities is parallelizable on GPU.
\end{itemize}

\IncMargin{0.5em}
\begin{algorithm}[t]
	\caption{V-cycle in multigrid solver}\label{alg:multigrid}
	\SetCommentSty{mycommfont}
	\SetKwInOut{AlgoInput}{Input}
	\SetKwInOut{AlgoOutput}{Output}
	\For{$l=0,\cdots,L-1$}
	{
		\If{$l>0$}{$\mathbf{T}^{l}\gets 0$}
		Relax $\mathbf{K}^{l}\mathbf{T}^{l}=\mathbf{f}^{l}$;\tcp*[f]{Relaxation}\\
		$\mathbf{r}^{l}=\mathbf{f}^{l}-\mathbf{K}^{l}\mathbf{T}^{l}$;\tcp*[f]{Residual update}\\
		$\mathbf{f}^{l+1}=R^{l+1}_{l}\mathbf{r}^{l}$;\tcp*[f]{Restrict residual}\\
	}
	Solve $\mathbf{K}^{L}\mathbf{T}^{L}=\mathbf{f}^{L}$ directly;\tcp*[f]{Solve on coarsest level}\\
	\For(\tcp*[f]{Go up in the V-cycle}){$l=L-1,\cdots,0$}
	{
		$\mathbf{T}^{l}\gets \mathbf{T}^{l}+I_{l+1}^{l}\mathbf{T}^{l+1}$;\tcp*[f]{Interpolate error \& correct}\\
		Relax $\mathbf{K}^{l}\mathbf{T}^{l}=\mathbf{f}^{l};$\tcp*[f]{Relaxation}\\
	}
\end{algorithm}
\DecMargin{0.5em}

\section{Implementation}\label{sec:implementation}

\begin{figure*}
	\centering
	\begin{overpic}[width=0.99\linewidth]{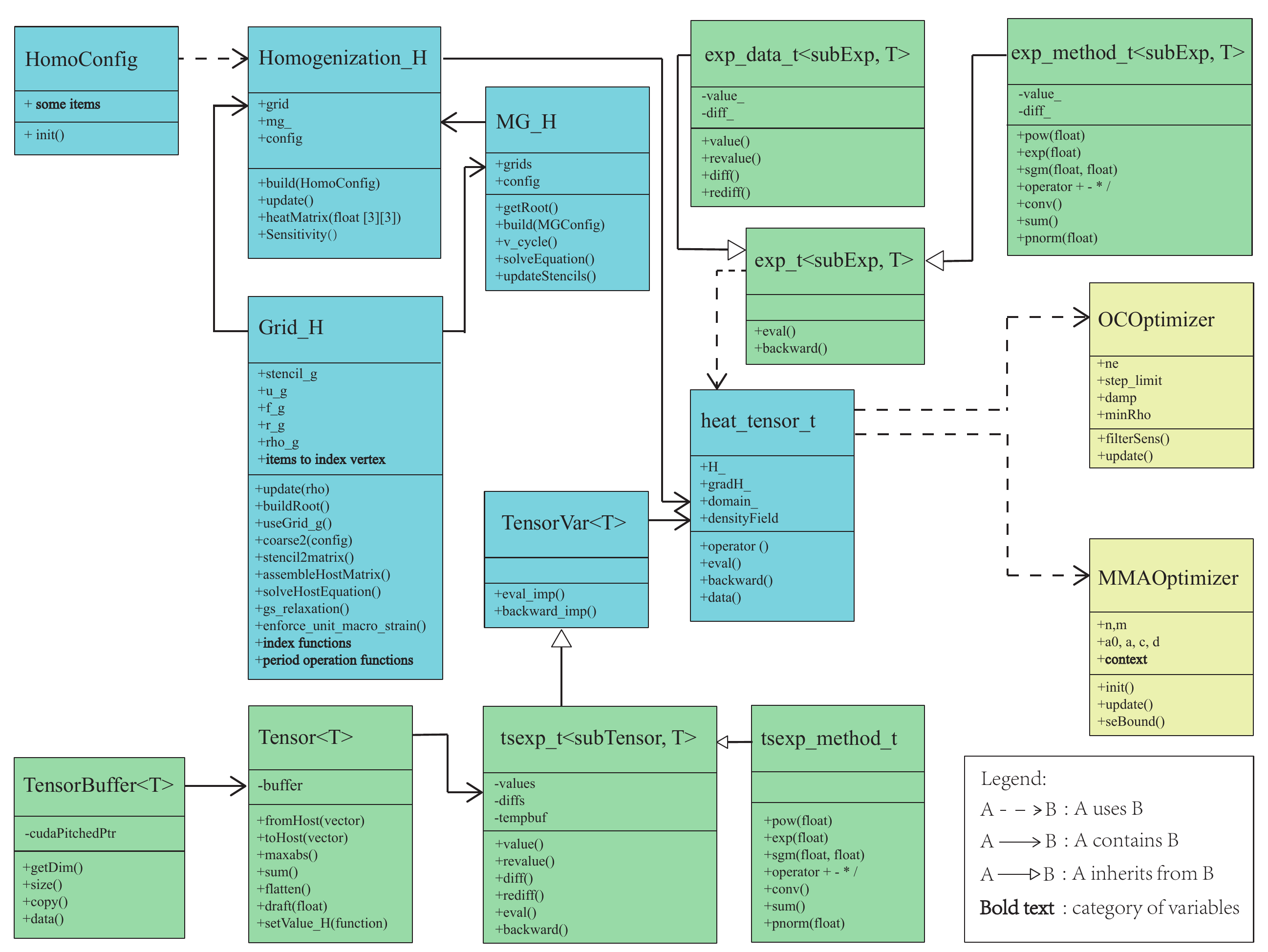}
		{
		}
	\end{overpic}
	\vspace{0mm}
	\caption{
		Hierarchy of the main classes
	}
	\label{fig:layers}
\end{figure*}

\subsection{A summary of the GPU implementation}
We inherited most of the techniques from \cite{zhang2023optimized}, such as the handling of periodic boundary conditions and the indexing method for adjacent elements, but we also made some modifications. Specifically, we chose to use simple floats for data storage to balance precision and computational speed rather than mixed representation. Since a mechanical node has strains in three directions, whereas a thermal node has only one temperature, we made corresponding adjustments to the data structure of the nodes.
\subsection{Setup and compiling}
Users can clone this framework or fork the current master branch from the GitHub repository: \url{https://github.com/quanyuchen2000/OPENTM}. 
The compilation and runtime environment mainly requires CUDA 11, Eigen3, glm, and OpenVDB. 
For installation, we also need Anaconda. 
We also provide a video to assist in the installation.
To extend the Python interface, we need pybind11~\cite{pybind11} and Anaconda. 
If users do not require extensive customization and obtain available microstructures with default optimization parameters, the Python interface described in Section~\ref{sec:e2u-interface} is suggested. 

\subsection{Code layout}
We elaborate on the classes in the code from bottom to top according to dependencies (Fig.~\ref{fig:layers}).

\paragraph{\lstinline{Tensor<T>}}
This class is designed to manage the density field on the GPU, utilizing a member buffer of type TensorBuffer. TensorBuffer handles memory allocation on the GPU and offers methods for interfacing with the host. 
Additionally, it supports basic vector operations, such as \lstinline{sum()} and \lstinline{flatten()}.

\paragraph{\lstinline{TensorVar<T>}}
This class inherits from \lstinline{tsexp_t}, derived from \lstinline{tsexp_method_t}. 
The \lstinline{tsexp_method_t} provides foundational methods for arithmetic operations and convolution, among others. 
The \lstinline{tsexp_t} class integrates \lstinline{Tensor<T>} to manage values and their derivatives, extending certain operations to access these values and derivatives. 
It also includes \lstinline{backward()} method for gradient computation. 
Together, these classes provide symbolic computation support on the GPU, streamlining the processing of density expressions and automatically calculating gradients.
Further details on usage can be found in Section~\ref{sec:compile-invoke}.

\paragraph{\lstinline{Grid_H}}
This class serves as the foundational layer, primarily managing essential attributes such as density, load, temperature variation, and the residual of each vertex on the GPU. 
It maintains the grid structure through functions tailored for indexing and periodic boundary conditions; due to their abundance, these functions are not enumerated in the hierarchy graph. 
The \lstinline{gs_relaxation} method implements an eight-color Gauss-Seidel relaxation technique for solving equations. 
Given that the coarsest layer is solved exactly, the \lstinline{assembleHostMatrix()} and \lstinline{solveHostEquation()} functions are employed to address the final layer.

\paragraph{\lstinline{MG_H}}
This class is designed for multi-grid operations, containing a vector of \lstinline{Grid_H} as its members. 
The \lstinline{solveEquation} function consolidates all the class methods to solve the equilibrium equation efficiently.

\paragraph{\lstinline{Homogenization_H}}
This class aggregates \lstinline{HomoConfig}, \lstinline{MG_H}, and \lstinline{Grid_H} to facilitate the homogenization operation. 
Its \lstinline{build()} function initiates by invoking the initialization functions of all its member variables. 
The \lstinline{update()} method refreshes the density values, while \lstinline{heatMatrix()} evaluates the equivalent thermal conductivity matrix. 
Additionally, \lstinline{Sensitivity()} provides the sensitivity analysis of the density variable.

\paragraph{\lstinline{heatTensor}}
This class, integrating \lstinline{Homogenization_H} and a density tensor, is engineered to model the thermal conductivity tensor as a function of density, in alignment with Eq.~\eqref{eq:homo}. 
The \lstinline{eval()} method computes the thermal conductivity tensor and its partial derivatives $\frac{\partial f}{\partial \kappa^H_{ij}}$. 
Additionally, the \lstinline{.backward(s)} function is designed for the effective backpropagation of gradients to the density variable. 
Users can define variables using the notation $(i, j)$ in a syntax similar to \lstinline{TensorVar<T>}. 
For instance, \lstinline{(Hh(0,0) - 0.5).pow(2) + (Hh(1,1) - 0.4).pow(2)} represents $(\kappa^H_{11} - 0.5)^2 + (\kappa^H_{22} - 0.4)^2$.

\paragraph{\lstinline{\\exp_t}}
This class, along with its subclasses, provides symbolic computation support for individual variables, akin to how \lstinline{TensorVar<T>} manages tensors. In our code, the type of \lstinline{Hh(i,j)} is \lstinline{\exp_t}.

\paragraph{\lstinline{OCOptimizer}}
This class implements the Optimality Criteria (OC) method and is utilized for filtering sensitivity and updating the density variable. 
The constructor prototype is \lstinline{OCOptimizer(float min_density, float stepLimit, float dampExponent)}. 
Here, \lstinline{min_density} specifies the minimum density threshold, while \lstinline{stepLimit} and \lstinline{dampExponent} define the maximum allowable step and damping ratio, respectively, for updating the density variable.

\paragraph{\lstinline{MMAOptimizer}}
This class implements the Method of Moving Asymptotes (MMA) and extends its application to GPU environments. 
The constructor is defined as \lstinline{MMAOptimizer(int nconstraint, int nvar, double a0_, double a_, double c_, double d_)}, where \lstinline{nconstraint} indicates the number of constraints and \lstinline{nvar} the number of variables. 
The parameters \lstinline{double a0_, double a_, double c_, double d_} correspond to those used in the original MMA method. The context includes several intermediate variables for computational support.

\subsection{Code invoking}\label{sec:compile-invoke}
After installing the framework, the process of designing 3D thermal microstructures includes the following steps:
(1) initialize parameters; (2) initialize homogenization; (3) define the material interpolation method; (4) create a thermal conductivity matrix; (5) define the objective function; (6) define the optimization process; (7) output the optimized result.
We provide a detailed illustration here.
\begin{enumerate}
	\item Initializing parameters. In the function \lstinline{HomoConfig::init()}, users set parameters including resolution, the way of density initialization, the volume ratio (not for constraint but for initialization), the heat ratio of two materials, and the target thermal conductivity tensor. 
	Since the tensor is symmetric, the parameter is a 1*6 vector.
	\item Initializing homogenization:
	\begin{lstlisting}
Homogenization hom_H(config);
TensorVar<float> rho_H(config.reso[0],config.reso[1],config.reso[2]);
initDensity(rho_H, config);
	\end{lstlisting}
	where \lstinline{config} is the defined above.
	\item Define the material interpolation method based on the SIMP approach~\citep{bendsoe1989optimal}:
	\begin{lstlisting}
auto rhop_H = rho_H.conv(radial_convker_t<float, Spline4>(1.5, 0)).pow(3) * (config.heatRatio[1] - config.heatRatio[0]) + config.heatRatio[1];
	\end{lstlisting}
	where \lstinline{conv(radial_convker_t<float, Spline4>(1.2))} means a convolution operation with the kernel \lstinline{radial_convker_t<float, Spline4>(1.5)} adopted in ~\citep{7332965}.
	\item Create a thermal conductivity matrix from the design domain \lstinline{hom_H} and the material interpolation method \lstinline{rhop_H}:%
	\begin{lstlisting}
heat_tensor_t <float, decltype(rhop_H)> Hh(hom_H, rhop_H);
	\end{lstlisting}
	\item Define the objective function $g(\bm{\kappa}^H, \bm{\kappa}^*)$, e.g., the function we used is $g(\bm{\kappa}^H, \bm{\kappa}^0) = ||\bm{\kappa}^H(:) - \bm{\kappa}^*(:)||^2$. The code can be written as:
	\begin{lstlisting}
auto objective = (Hh(0, 0) - tt[0]).pow(2) + (Hh(1, 1) - tt[1]).pow(2) +(Hh(2, 2) - tt[2]).pow(2) + (Hh(0, 1) - tt[3]).pow(2) + (Hh(2, 1) - tt[4]).pow(2) + (Hh(0, 2) - tt[5]).pow(2);
	\end{lstlisting}
	where \lstinline{tt} is the target tensor $\bm{\kappa}^*(:)$, \lstinline{Hh} is $\bm{\kappa}^H$. 
	\item Define the optimization process. We utilize the variable \textbf{mode} to switch between different optimization models. Considering both performance and code complexity, we adopt Model~\eqref{md:model3} as an illustrative example:
	\begin{lstlisting}
OCOptimizer oc(ne, 0.001, 0.02, 0.5);
VolumeGovernor governor;
for (int itn = 0; itn < config.max_iter; itn++) {
float val = objective.eval();
objective.backward(1);
float lowerBound = rhop_H.sum().eval_imp() / pow(reso, 3);
float volfrac = rho_H.sum().eval_imp() / pow(reso, 3);
if (governor.volume_check(val, lowerBound, volfrac, itn)) {
	printf("converged"); break;
}
if (criteria.is_converge(itn, val) && governor.get_current_decrease() < 1e-4) 
{ printf("converged\n"); break; }
auto sens = rho_H.diff().flatten();
auto rhoarray = rho_H.value().flatten();
int ereso[3] = { reso,reso,reso };
oc.filterSens(sens.data(), rhoarray.data(), reso, ereso);
oc.update(sens.data(), rhoarray.data(), governor.get_volume_bound());
rho_H.value().graft(rhoarray.data());
}
	\end{lstlisting}
	In this process, we first compute both the value and gradient of the objective in each iteration. Subsequently, the volume fraction boundary is recalculated, enabling the governor to dynamically modify this boundary. The derived data is then used to execute a step in the optimization process and to update the density accordingly.
	\item Output the optimized density field and elastic matrix:
	\begin{lstlisting}
rho.value().toVdb(getPath("rhoFile"));
Ch.writeToTxt(getPath("ChFile"));
	\end{lstlisting}
	where \lstinline{getPath} is a function to prefix the output directory to a given string.
	The member function \lstinline{toVdb} writes the data of \lstinline{TensorVar} to an OpenVDB file.
\end{enumerate}

Users only need to define \lstinline{config}, the material interpolation method, and the objective function before running the code to solve the IHPs.
%
The outputs contain
a microstructure visualization file ( \lstinline{*.vbd}), an elastic tensor matrix (\lstinline{*.txt}),
Users can use Rhino to visualize 
\lstinline{*.vdb} files.

\subsection{Easy-to-use interfaces}\label{sec:e2u-interface}

\paragraph{Environment} 
Ensure that CUDA is installed and the PATH is set correctly. The Python library (.pyd files) is located in the ‘python' folder. All required libraries (.dll files) are packaged in the same folder.

\paragraph{Python interface} 
The Python interface can be used as:
\lstset{
	keywordstyle=\color{orange},
	language = python,
}
\begin{lstlisting}
import homo3d
import numpy as np
help(homo3d)
rho = homo3d.runInstance(32, [1, 0.0001], [0.5, 0.4, 0.3, 0, 0, 0], homo3d.InitWay.IWP, homo3d.Model.oc)
matrix_3d = np.reshape(rho, (32, 32, 32))
print(matrix_3d)
\end{lstlisting}
The command \lstinline{help(homo3d)} helps users understand the available functions and classes. The function \lstinline{runInstance} is used to generate the structure with parameters including resolution, thermal coefficients for two materials, the target thermal tensor, initialization method, and optimization model. The density is saved to a \lstinline{.vdb} file in the specified folder and is also returned as rho. This data can be reshaped using \lstinline{np.reshape} to obtain voxel data for visualization in Python.

We also have another method to call the interface:
\begin{lstlisting}
import homo3d
import numpy as np
	
help(homo3d)
reso = 16
homogenization = homo3d.homo()
homogenization.setConfig(reso, [1., 0.0001], [0.5, 0.4, 0.3, 0., 0., 0.])
	
# initialize 32*32*32 array and flatten it
flattened_array = [0.5]*(32*32*32)
	
homogenization.setDensity(flattened_array)
rho = homogenization.optimize()
print(rho)
\end{lstlisting}
where user can define their own initialization. Meanwhile, this encapsulation method is more conducive to maintenance and expansion.


\paragraph{Compile your own library} 
We provide the compilation code of the interface in the main function as comments, which helps avoid excessive dependencies in the project. To invoke these codes to compile the library, users must uncomment the corresponding code, configure the corresponding dependencies, and then compile.
\section{Examples}\label{sec:example}
\paragraph{Optimization parameters}
The material penalization factor $p$ is 3, the filter radius $r$ is 1.5, the maximum iteration number $I_{\max}$ is 500, the step size of density is 0.05, and the damping factor of OC method is 0.5. 
The cubic domain is discretized with 8-node brick elements. 
The default resolution is $128\times128\times128$, and the default initialization is IWP (Schoen's I-graph–wrapped package). 
The thermal conductivity coefficient of solids is $\kappa_0 = 1$ and of voids is $\kappa_{min} = 10^{-4}$. Unless otherwise specified, our experiments are conducted on a computer with an RTX 2060 graphics card.

\begin{figure}[t]
	\centering
	\begin{overpic}[width=0.99\linewidth]{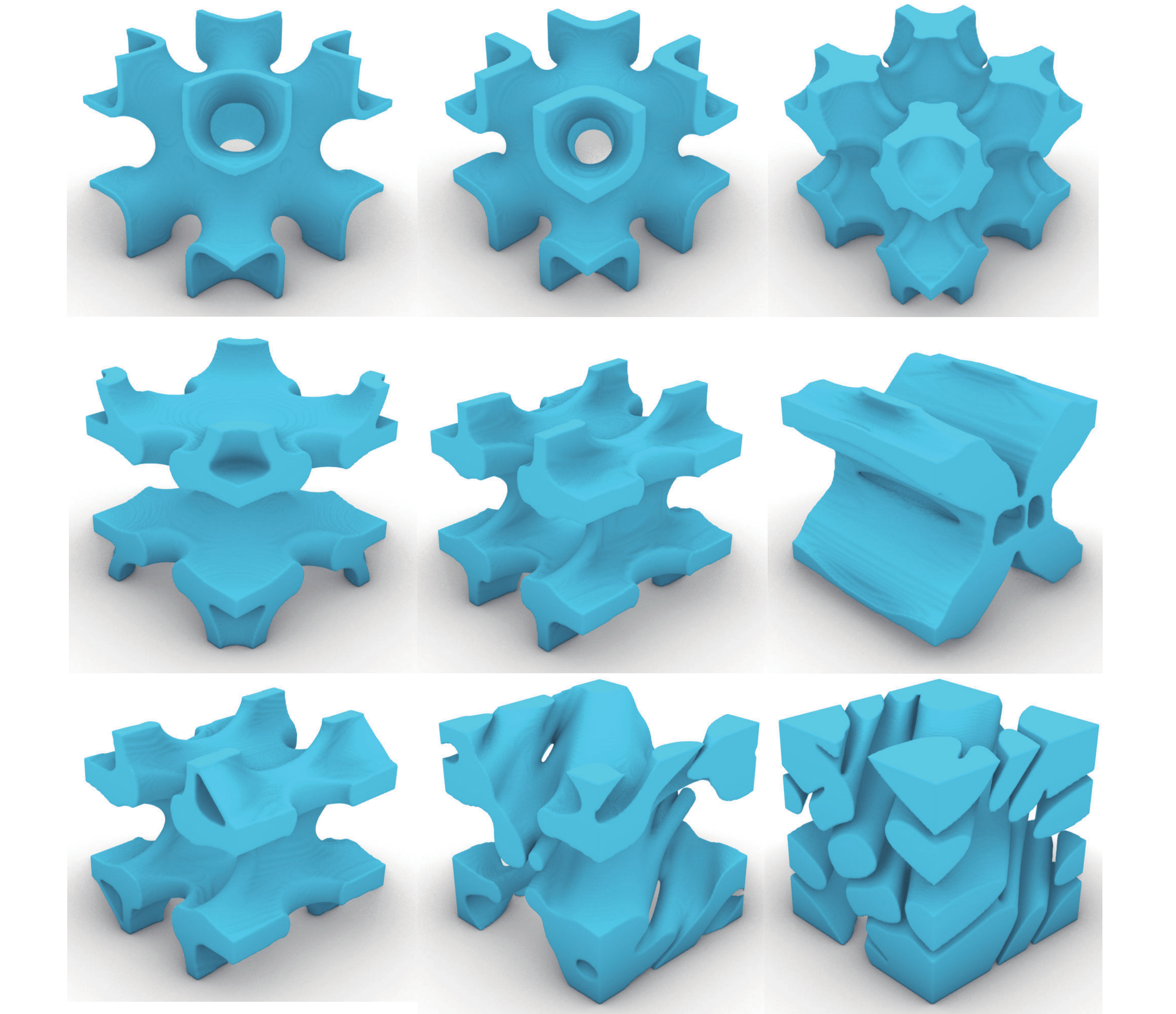}
		{
			\put(0,25){ (c)}
			\put(0,55){ (b)}
			\put(0,85){ (a)}
		}
	\end{overpic}
	\vspace{-2mm}
	\caption{
		Various types of target thermal conductivity tensors: isotropy (a), orthotropic anisotropy (b), and general case (c).
	}
	\label{fig:customized}
\end{figure}

\begin{figure}[t]
	\centering
	\begin{overpic}[width=0.99\linewidth]{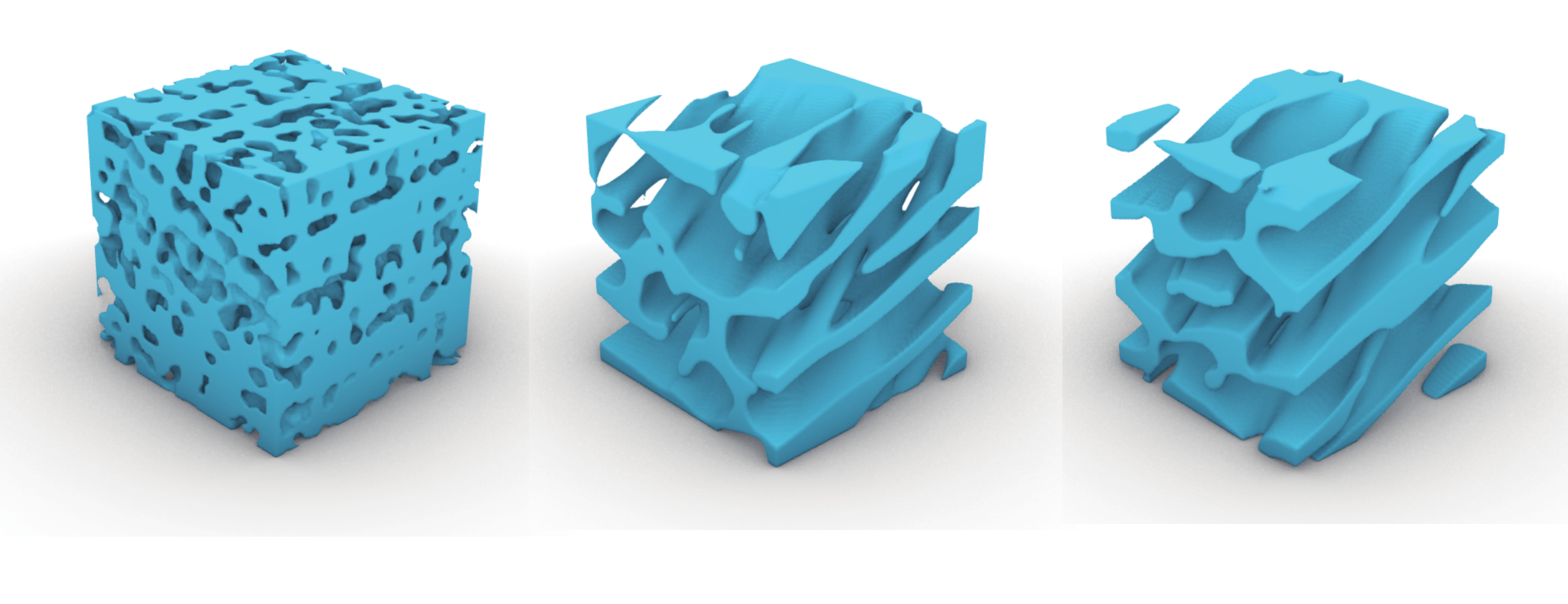}
		{
			\put(72,0.5){ With symmetry }
			\put(37,0.5){ Without symmetry}
			\put(8,0.5){ Initialization}
		}
	\end{overpic}
	\vspace{-1mm}
	\caption{ With/without central symmetry operations under an asymmetric random initialization with target tensor $[0.3, 0.2, 0.1, 0.05, 0.01, 0.1]$. 
	}
	\label{fig:symmetric_result}
\end{figure}

\begin{figure}[t]
	\centering
	\begin{overpic}[width=1\linewidth]{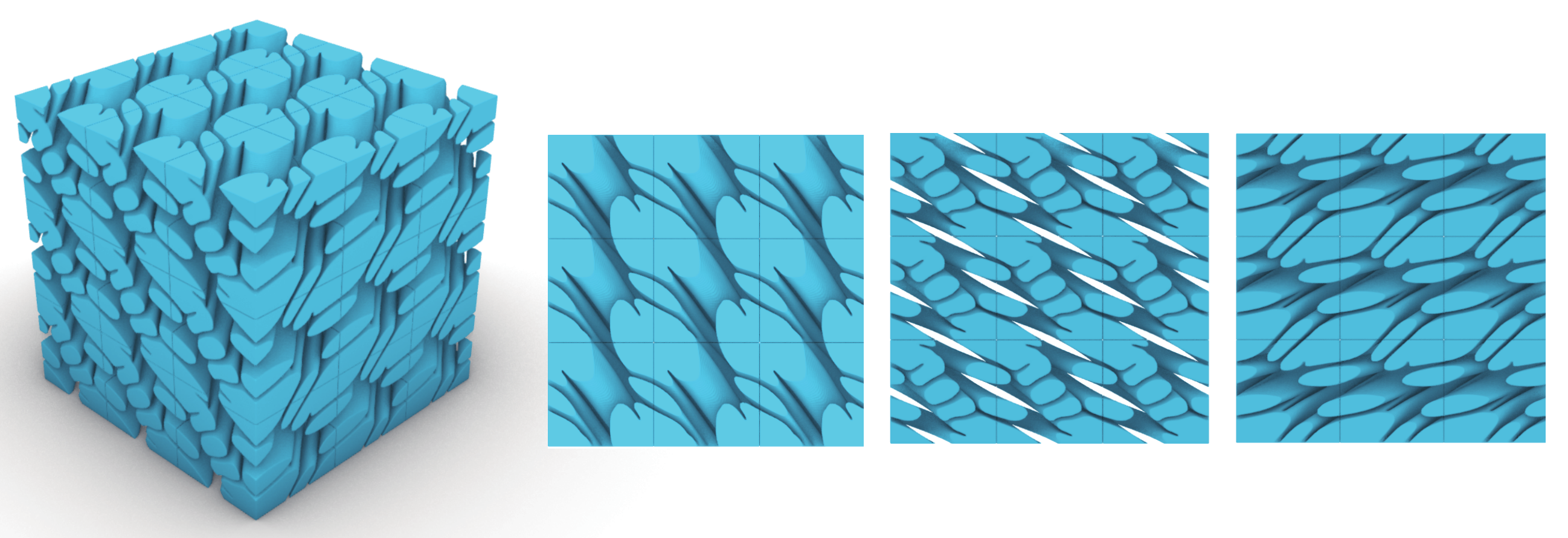}
		{
			\put(43, 0){top}
			\put(63, 0) {front}
			\put(85, 0) {right}
		}
	\end{overpic}
	\vspace{-3mm}
	\caption{
		A thermal metamaterial with $3\times3\times3$ arrangement of the microstructure in Fig.~\ref{fig:customized}~(c) right and its three perspectives.
	}
	\label{fig:3view}
\end{figure}

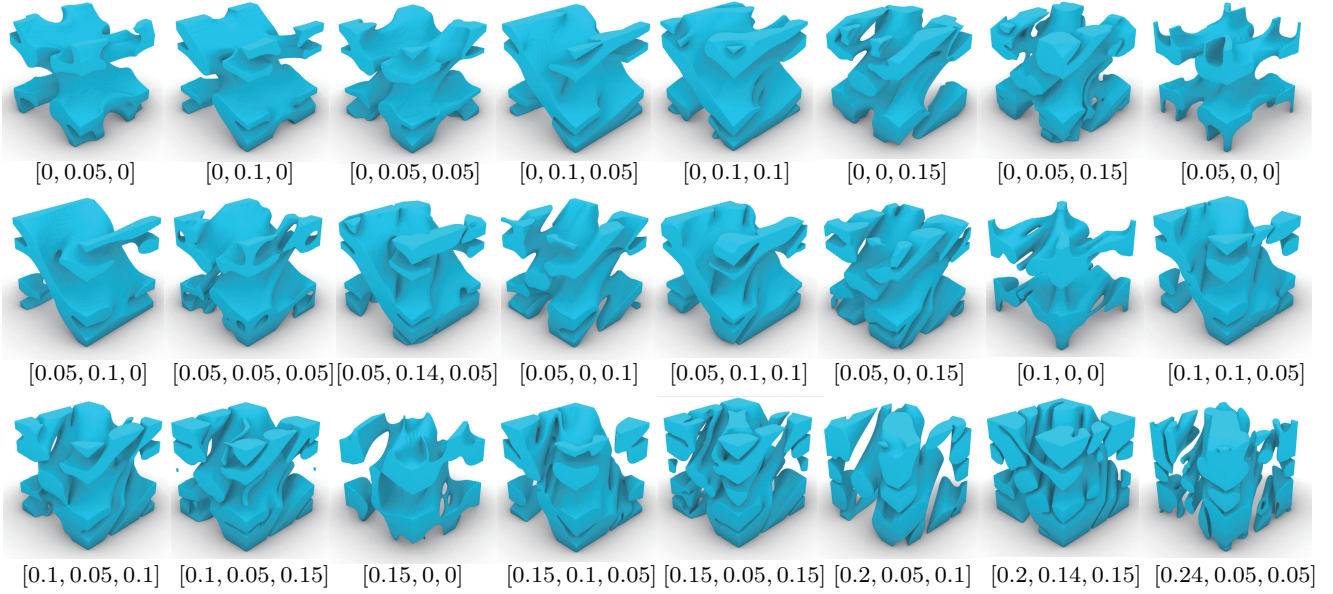
\begin{figure*}[!t]
	\centering
	\begin{overpic}[width=0.99\linewidth]{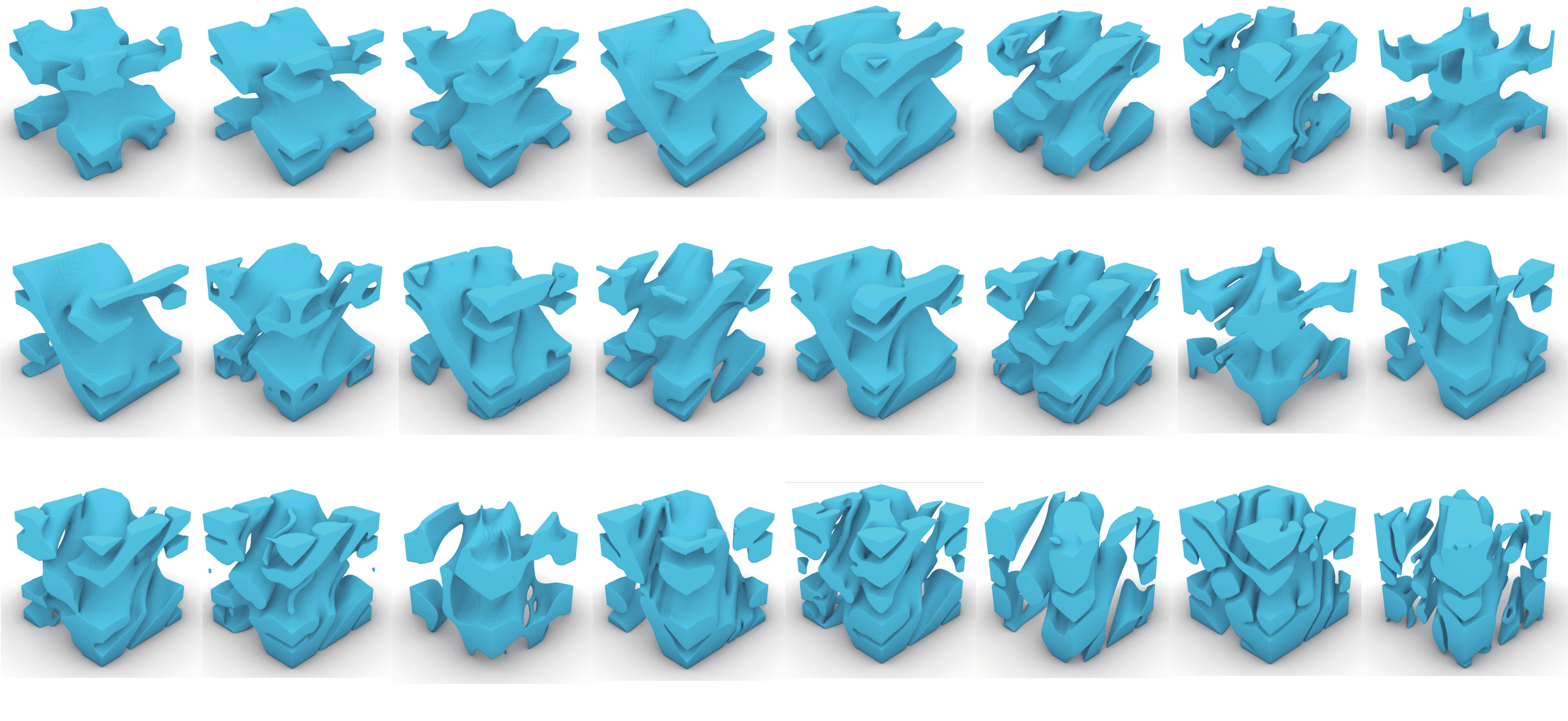}
		{
			\put(2,32){ $[0, 0.05, 0]$}
			\put(1.5,16.5){ $[0.05, 0.1, 0]$}
			\put(1,1){ $[0.1, 0.05, 0.1]$}
			\put(15,32){ $[0, 0.1, 0]$}
			\put(12.5,16.5){ $[0.05, 0.05, 0.05]$}
			\put(13,1){ $[0.1, 0.05, 0.15]$}
			\put(26,32){ $[0, 0.05, 0.05]$}
			\put(25,16.5){ $[0.05, 0.14, 0.05]$}
			\put(27,1){ $[0.15, 0, 0]$}
			\put(39,32){ $[0, 0.1, 0.05]$}
			\put(39,16.5){ $[0.05, 0, 0.1]$}
			\put(38,1){ $[0.15, 0.1, 0.05]$}
			\put(51.5,32){ $[0, 0.1, 0.1]$}
			\put(50.5,16.5){ $[0.05, 0.1, 0.1]$}
			\put(50,1){ $[0.15, 0.05, 0.15]$}
			\put(64,32){ $[0, 0, 0.15]$}
			\put(63,16.5){ $[0.05, 0, 0.15]$}
			\put(63,1){ $[0.2, 0.05, 0.1]$}
			\put(75.5,32){ $[0, 0.05, 0.15]$}
			\put(77,16.5){ $[0.1, 0, 0]$}
			\put(75,1){ $[0.2, 0.14, 0.15]$}
			\put(89.5,32){ $[0.05, 0, 0]$}
			\put(88.5,16.5){ $[0.1, 0.1, 0.05]$}
			\put(87.5,1){ $[0.24, 0.05, 0.05]$}
		}
	\end{overpic}
	\vspace{-2mm}
	\caption{
		Gallery of our optimized microstructures by traversing the feasible tensor space. We fix the diagonal elements by [$\kappa_{11}^H$, $\kappa_{22}^H$, $\kappa_{33}^H$] = $[0.3, 0.2, 0.1]$ and below each image are their [$\kappa_{12}^H$, $\kappa_{23}^H$, $\kappa_{13}^H$].
	}
	\label{fig:gallary}
\end{figure*}
\subsection{Customized thermal microstructures}
In addition to specifying the thermal conductivity tensor, the symmetry, isotropy, and other properties of thermal materials can also be constrained by tensor matrices. 
\begin{itemize}
	\item Isotropic materials exhibit identical properties in all directions. For isotropic materials, we have $\kappa_{11}^H = \kappa_{22}^H = \kappa_{33}^H$ and $\kappa_{ij}^H = 0$ (for $i \neq j$), resulting in only one unique parameter. Fig.~\ref{fig:customized}~(a) illustrates three isotropic microstructures, where $\kappa_{ii}^H$ equals 0.1, 0.2, and 0.3 from left to right, respectively. These generated structures retain the features of the initial IWP structure, while the volume fraction increases with higher $\kappa_{ii}^H$. 
	\item Orthotropic anisotropic materials exhibit varying properties along different axes. For orthotropic anisotropy, we have $\kappa_{ij}^H=0$ (for $i \neq j$), with no constraints imposed on $\kappa_{ii}^H$. In Fig.~\ref{fig:customized}~(b), from left to right, the $[\kappa_{11}^H,\kappa_{22}^H,\kappa_{33}^H]$ are depicted as $[0.3,0.3,0.1]$, $[0.3,0.2,0.1]$, and $[0.3,0.5,0.1]$, respectively. 
	\item Central symmetry indicates that for any point on the object, there is another point directly opposite it at the same distance from the center. To achieve this constraint, in each iteration, we identify two elements at symmetrical positions and average their densities and the derivatives of the objective function with respect to their densities, then assign this average value. 
	Fig.~\ref{fig:symmetric_result} shows results with and without central symmetry constraints.
	\item General case. 
	Due to the excessive number of free variables in this case, we fix the diagonal elements by $[\kappa_{11}^H,\kappa_{22}^H,\kappa_{33}^H]=[0.3,0.2,0.1]$. In Fig.~\ref{fig:customized}~(c), from left to right, $[\kappa_{12}^H,\kappa_{23}^H,\kappa_{13}^H]$ takes the values $[0.01,0.01,0]$, $[0.1,0.05,0.05]$, and $[0.25,0.17,0.14]$, respectively. 
	On the left side, the situation arises when the cross terms are relatively small, while the middle depicts moderate cross terms, and the right side illustrates the limits of these cross terms. Given the constraints imposed by the second law of thermodynamics, mandating the thermal conductivity tensor to be positive definite, the scenario on the right showcases a near-zero determinant. 
	Fig.~\ref{fig:3view} shows the structure of the right side of Fig.~\ref{fig:customized}~(c) assembled by $3\times3\times3$ arrangement to explain how structures that appear very fragmented and scattered in a single packet are seamlessly combined together in an array.
\end{itemize}

\subsection{Traversing anisotropic space}
\cite{sha2022topology} perform an appealing work to traverse full-parameter anisotropic space in 2D. We have extended this work to 3D and attempted to observe the traversal results. 
We systematically explore the feasible tensor space by traversing it in increments of 0.05, with the diagonal elements held constant. 
Since the determinant of the principal minors must be greater than 0, this requirement ensures that $|\kappa_{12}^H|<\sqrt{0.3\times0.2}$, $|\kappa_{13}^H|<\sqrt{0.3\times0.1}$, $|\kappa_{23}^H|<\sqrt{0.2\times0.1}$. Assuming $\kappa_{ij}^H > 0$, the possible values for 
$\kappa_{12}^H$ are in $\{0, 0.05, 0.1, 0.15, 0.2, 0.24\}$, for $\kappa_{23}$ are in $\{0, 0.05, 0.1, 0.15\}$, and for 
$\kappa_{13}^H$ are in $\{0, 0.05, 0.1, 0.14\}$. 
This results in a total of 96 combinations. 
After excluding those that produce a negative determinant, 61 viable test cases remain. 

\begin{figure}[t]
	\centering
	\begin{overpic}[width=0.99\linewidth]{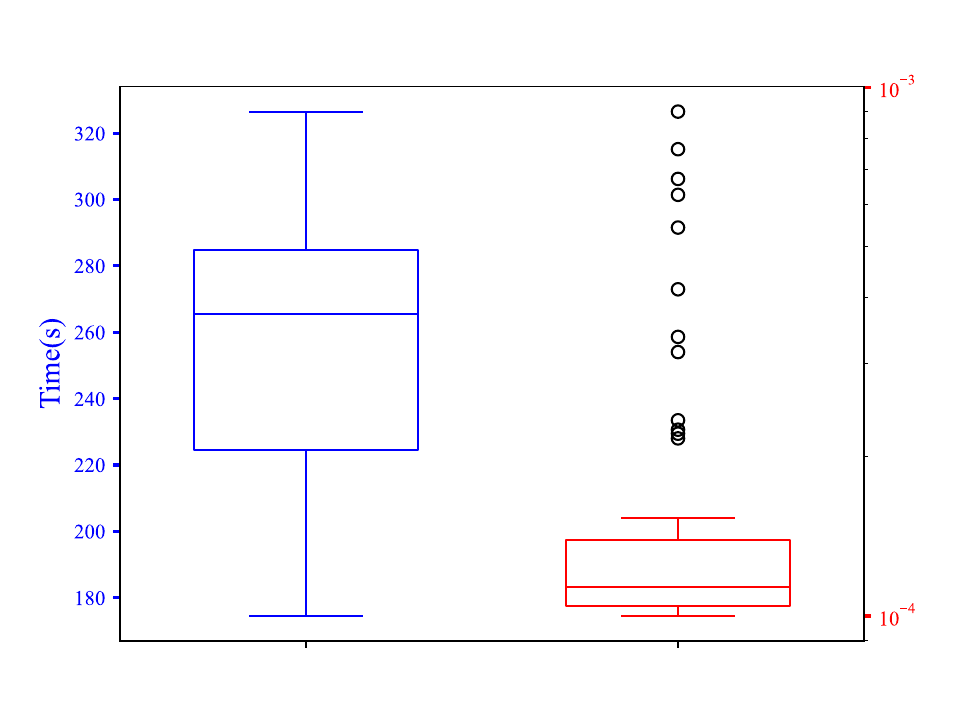}
		{
			\put(23,3){ Solve time }
			\put(61,3){ $g(\kappa^H,\kappa^0)$ }
		}
	\end{overpic}
	\vspace{-3mm}
	\caption{
		The statistical results (calculation time vs. the solution error). 
		The maximum tolerable error is set to $10^{-4}$. 
	}
	\label{fig:gallary static}
\end{figure}

Fig.~\ref{fig:gallary} presents a selection of our generated results alongside their respective tensors, while Fig.~\ref{fig:gallary static} showcases the statistical outcomes of these solutions. This illustrates that our framework is capable of delivering reasonably accurate results within a practical time.


\begin{figure}[t]
	\centering
	\begin{overpic}[width=0.95\linewidth]{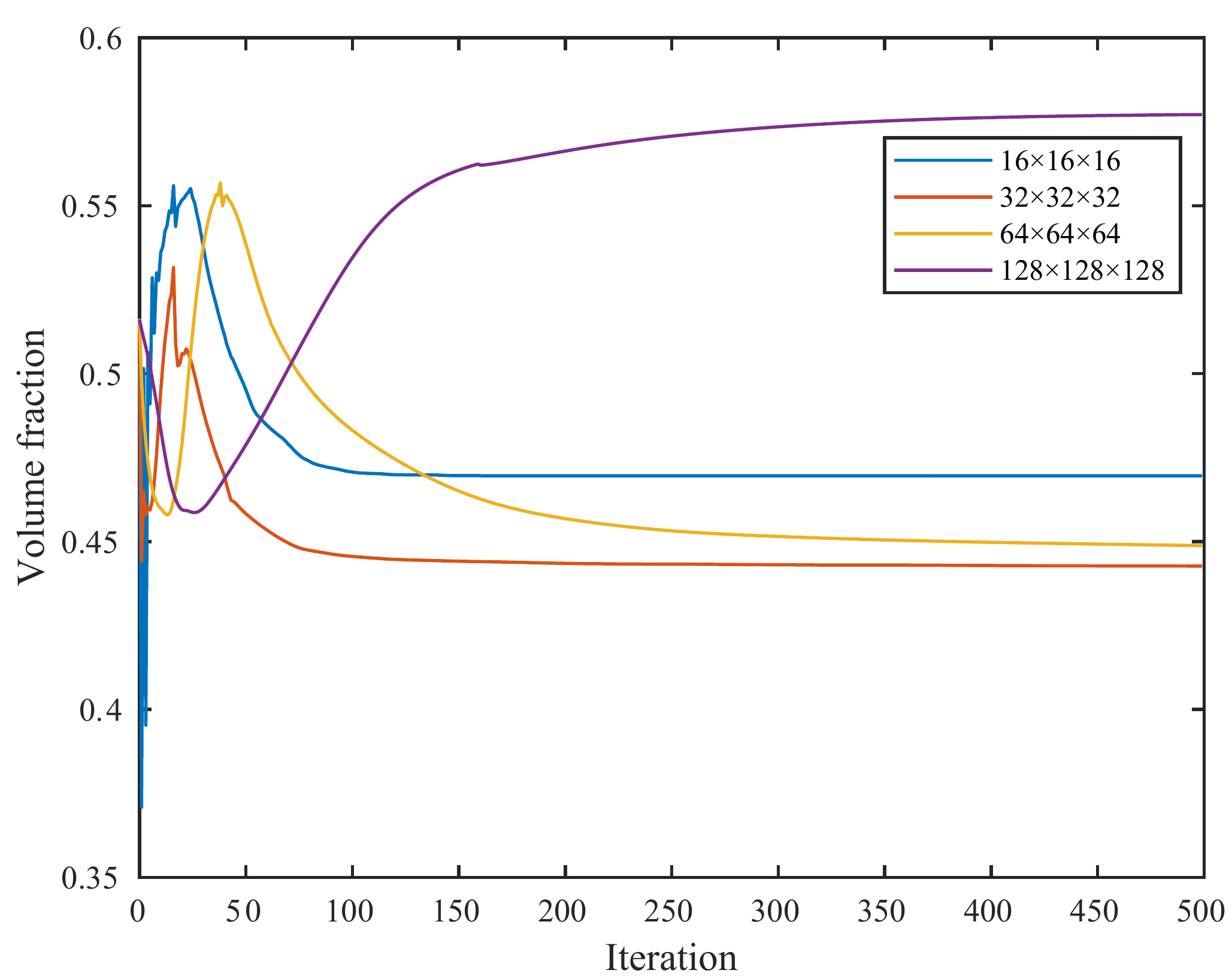}
		{
		}
	\end{overpic}
	\vspace{0mm}
	\caption{
		Solving model~\eqref{md:model2} by MMA with different resolution.
	}
	\label{fig:mma}
\end{figure}

\subsection{Parameter discussions}
\subsubsection{Different solvers}
\begin{figure}[t]
	\centering
	\begin{overpic}[width=1.05\linewidth]{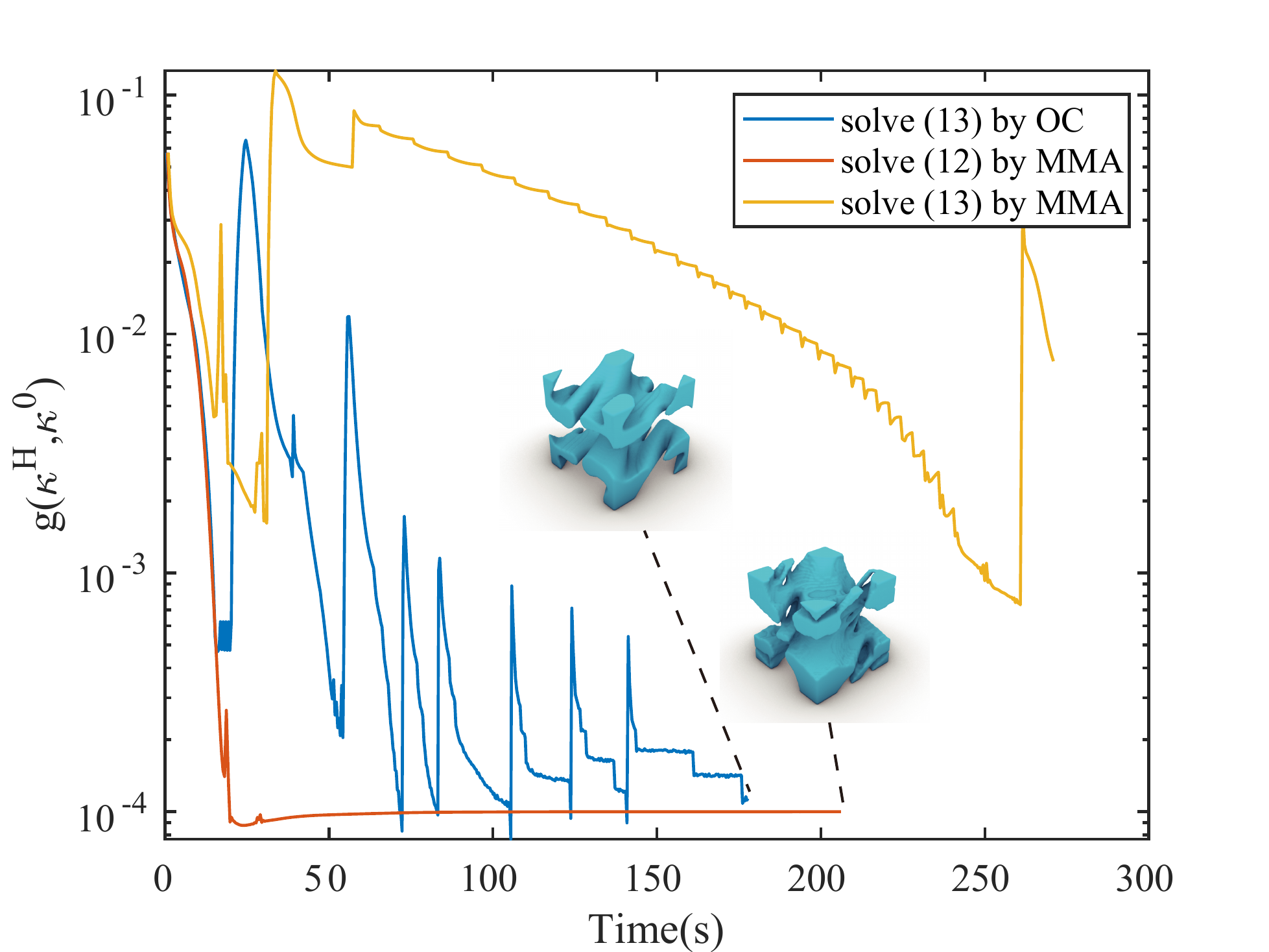}
		{
		}
	\end{overpic}
	\vspace{0mm}
	\caption{
		Comparison of iteration curves for solving the optimization model~\eqref{md:model3} by MMA and OC, and solving model~\eqref{md:model2} by MMA at a resolution of $64\times 64\times 64$. The target tensor is [0.3, 0.2, 0.1, 0.1, 0.05, 0.05]. Solving model~\eqref{md:model3} by MMA does not converge in 500 iterations.
	}
	\label{fig:mmaVsoc}
\end{figure}

To gain a deeper understanding of the effects of different solvers, we discuss the MMA and OC algorithms. Since model~\eqref{md:model2} cannot be solved by OC, we compare model~\eqref{md:model2} solved by MMA and model~\eqref{md:model3} solved by MMA and OC in Fig.\ref{fig:mmaVsoc}. Obviously, model~\eqref{md:model2} is more suitable for MMA. Thus, in our work, the comparison is between the optimization model~\eqref{md:model2} solved by MMA and the optimization model~\eqref{md:model3} optimized by OC. 
We compare and analyze these two solvers (MMA vs. OC) from three aspects: resolution, memory usage, and time consumption. 

\begin{figure*}[t]
	\centering
	\begin{overpic}[width=0.95\linewidth]{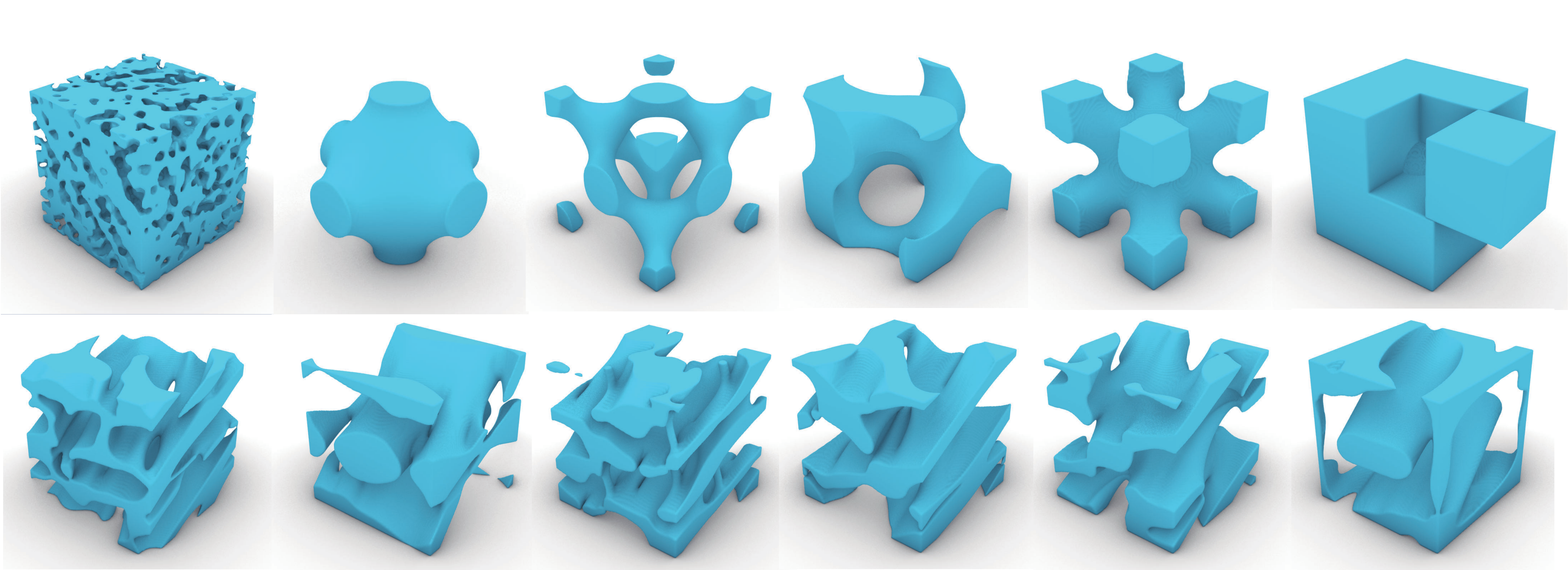}
		{
			\put(3,34){ random input}
			\put(24,34){ P }
			\put(41,34){ D}
			\put(56,34){ G}
			\put(71,34){ IWP}
			\put(86,34){ center ball}
		}
	\end{overpic}
	\vspace{-1mm}
	\caption{
		We show different initial density fields (upper row) and optimized results (bottom row) for the same target tensor. From left to right, these initial structures are sequentially random center, P, D, G, IWP, and center ball. 
		The random center is the random initialization proposed by~\cite{zhang2023optimized}. 
		P, D, G, and IWP are classic open-source models. 
		A center ball is to cut a small sphere out from the center of a cube with a certain volume fraction.
		\cite{sigmund200199} used it in Matlab using 88 lines of code. The target tensor is [0.3 0.2 0.1 0.05 0 0.1].
	}
	\label{fig:initialization}
\end{figure*}

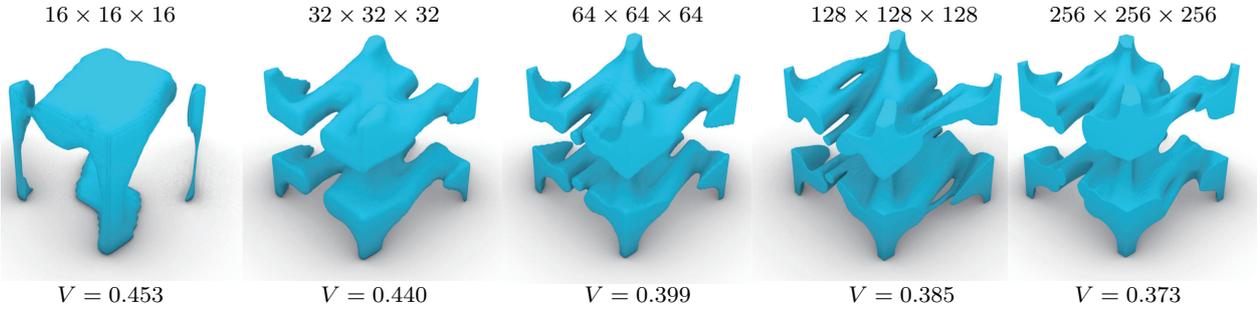
\begin{figure*}[t]
	\centering
	\begin{overpic}[width=0.95\linewidth]{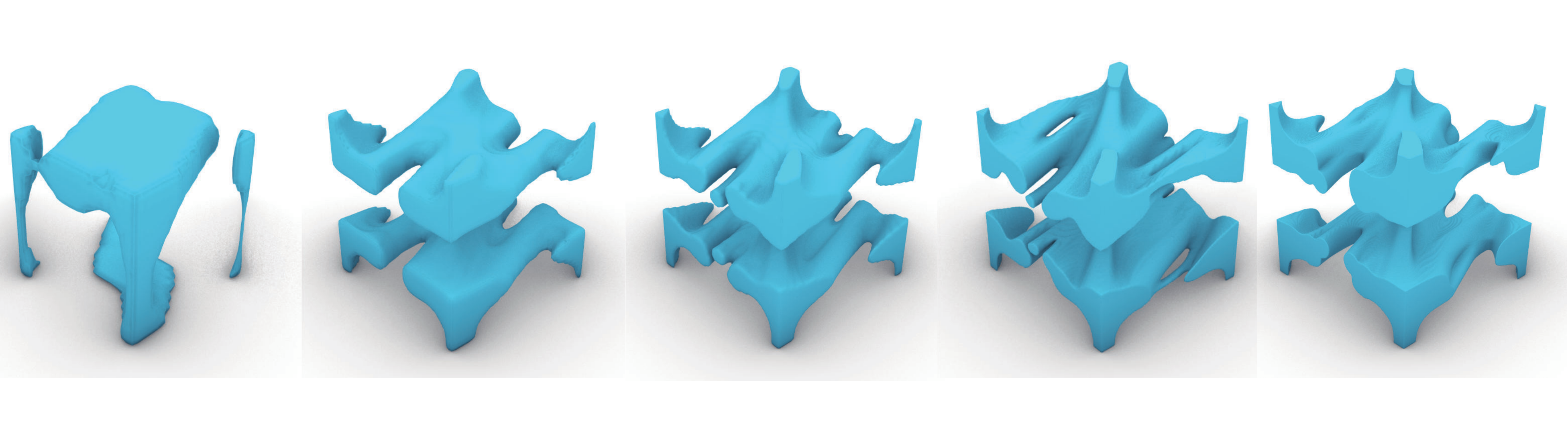}
		{
			\put(3,24){ $16\times16\times16$}
			\put(4,1.5){ $V = 0.453$}
			\put(25,1.5){ $V = 0.440$}
			\put(24,24){ $32\times32\times32$}
			\put(46,1.5){ $V = 0.399$}
			\put(45,24){ $64\times64\times64$}
			\put(67,1.5){ $V = 0.385$}
			\put(64,24){ $128\times128\times128$}
			\put(85,1.5){ $V = 0.373$}
			\put(83,24){ $256\times256\times256$}
		}
	\end{overpic}
	\vspace{-3mm}
	\caption{
		Volume fraction v.s. resolution. We kept all conditions constant except for the resolution. The resolution is displayed at the top of the image, and the corresponding volume fraction results are shown at the bottom. The target tensor is [0.3, 0.2, 0.1, 0.1, 0, 0].
	}
	\label{fig:reso vs volume}
\end{figure*}

\begin{itemize}
	\item Resolution. 
	We tested the MMA on voxel models with resolutions of $16^3$, $32^3$, $64^3$, and $128^3$ for solving~\eqref{md:model2}, achieving $g$ below $10^{-4}$ in all cases. Iteration curves with different resolutions are shown in Fig.~\ref{fig:mma}. Except for the $128^3$ resolution case, the volume fraction initially increases during the optimization process and then decreases. This occurs because, at the start, the optimizer prioritizes achieving the target values for the thermal conductivity tensor, as specified by the constraint in \eqref{md:model2}. Once these target values are met, the optimizer gradually reduces the volume fraction. 
	In the case of a $128^3$ resolution, although the thermal conductivity tensor reaches the target values, the optimization process cannot converge to the optimal solution due to numerical limitations. 
	For $128^3$ case, $\frac{\partial V}{\partial \rho} = 1/(128^3) \approx 4.8\times10^{-7}$ near the accuracy of float. Given the precision limits of floating-point arithmetic and the complex calculations involved in the MMA method, the results are particularly susceptible to numerical error.
	\item Memory usage. To optimize a structure with a resolution of $128^3$, OC uses 348MB of GPU memory, whereas MMA uses 1312MB. The additional memory in MMA is required to store the necessary intermediate variables. To facilitate parallelism, many of these intermediate variables are set to the same size as the resolution, resulting in substantial memory overhead for intermediate computations.
	
	\item Time consuming. 
	Since MMA does not support a resolution of $128^3$ or higher resolution, we compare MMA and OC at a resolution of $64^3$ to solve the optimization model~\eqref{md:model3}. 
	As shown in Fig.~\ref{fig:mmaVsoc}, OC completes iterations slightly earlier than MMA.
	For MMA solver, $g$ quickly drops below the maximum limit and then gradually recovers. 
	In contrast, for the OC solver, $g$ reaches and exceeds the limit, and a rebound occurs due to changes in the volume constraint, leading to an oscillatory convergence near the maximum limit.
\end{itemize}

In summary, MMA is limited by memory capacity and numerical precision, preventing it from solving large-scale problems. 
However, as far as it can manage, MMA achieves a higher degree of constraint satisfaction during its solution process. 
Additionally, the volume fraction consistently decreases monotonically, a feature not present in the OC method.

\subsubsection{Density initialization}
We inherit the random initialization method of \cite{zhang2023optimized} and add some classical initializations. 
The results obtained by solving the model~\eqref{md:model3} are shown in Fig.~\ref{fig:initialization}. 
The material distribution of the result is affected by the initialization. 

\subsubsection{Relation between resolution and volume fraction}

With increasing resolution, the design space expands, yielding structures with finer details. Therefore, to achieve a consistent thermal conductivity tensor, structures generated by solving the model~\eqref{md:model3} at higher resolutions are expected to exhibit smaller volume fractions. We conducted experiments at resolutions of $16^3, 32^3, 64^3, 128^3,$ and $256^3$ under identical conditions. As depicted in Fig.~\ref{fig:reso vs volume}, the results align with these anticipations.

\subsection{Extensions}
\label{sub:ext}

\paragraph{Different Objectives}
Our framework uses automatic differentiation (AD) techniques to facilitate seamless extension of our program to optimize various objectives. Users can effortlessly modify the code to alter the formulation of the objectives without the need for redundant calculations. The default objective function employed in our framework is:

\begin{small}
	\begin{equation}
		\label{eq:mse}
		\begin{split}
			g(\bm{\kappa}^H, \bm{\kappa}^*)= &
			(\kappa^H_{11} - \kappa^*_{11})^2+(\kappa^H_{22} - \kappa^*_{22})^2+(\kappa^H_{33} - \kappa^*_{33})^2 \\
			&  +(\kappa^H_{12} - \kappa^*_{12})^2 + (\kappa^H_{23} - \kappa^*_{23})^2 + (\kappa^H_{13} - \kappa^*_{13})^2,
		\end{split}
	\end{equation}
\end{small}
where $\bm{\kappa}^*$ is target tensor. The corresponding code is: 
\begin{lstlisting}
auto objective = (Hh(0, 0)-tt[0]).pow(2) + (Hh(1, 1)-tt[1]).pow(2) + (Hh(2, 2)-tt[2]).pow(2) + (Hh(0, 1)-tt[3]).pow(2) + (Hh(2, 1)-tt[4]).pow(2) + (Hh(0, 2)-tt[5]).pow(2);
\end{lstlisting}


To place greater emphasis on the values of the cross terms, we can design the objective function as follows:
\begin{small}
	\begin{equation}
		\label{eq:fracmse}
		\begin{split}
			g(\bm{\kappa}^H, \bm{\kappa}^*)&=
			(\kappa^H_{11}/\kappa^*_{11}-1)^2+(\kappa^H_{22}/\kappa^*_{22}-1)^2+(\kappa^H_{33}/\kappa^*_{33}-1)^2\\
			& +(\kappa^H_{12}/\kappa^*_{12}-1)^2 + (\kappa^H_{23}/\kappa^*_{23}-1)^2 + (\kappa^H_{13}/\kappa^*_{13}-1)^2.
		\end{split}
	\end{equation}
\end{small}
Similarly, the corresponding code is:
\begin{lstlisting}
auto objective = (Hh(0, 0)/tt[0]-1).pow(2) + (Hh(1, 1)/tt[1]-1).pow(2) + (Hh(2, 2)/tt[2]-1).pow(2) + (Hh(0, 1)/tt[3]-1).pow(2) + (Hh(2, 1)/tt[4]-1).pow(2) + (Hh(0, 2)/tt[5]-1).pow(2);
\end{lstlisting}
Compared to~\eqref{eq:mse}, the cross-term error of~\eqref{eq:fracmse}  is smaller.
In fact, optimizing~\eqref{eq:mse} distributes the error equally among all components by value, while minimizing~\eqref{eq:fracmse} distributes the error in equal proportions to each component.

Although the $\ell_1$-norm is non-differentiable, it has many applications in optimization. Here, we provide an example of defining an objective using the $\ell_1$-norm:
\begin{small}
	\begin{equation}
		\label{eq:l1}
		\begin{split}
			g(\bm{\kappa}^H, \bm{\kappa}^*)= &
			|\kappa^H_{11} - \kappa^*_{11}|+|\kappa^H_{22} - \kappa^*_{22}|+|\kappa^H_{33} - \kappa^*_{33}| \\
			&  +|\kappa^H_{12} - \kappa^*_{12}| + |\kappa^H_{23} - \kappa^*_{23}| + |\kappa^H_{13} - \kappa^*_{13}|. 
		\end{split}
	\end{equation}
\end{small}

\begin{figure}[t]
	\centering
	\begin{overpic}[width=0.99\linewidth]{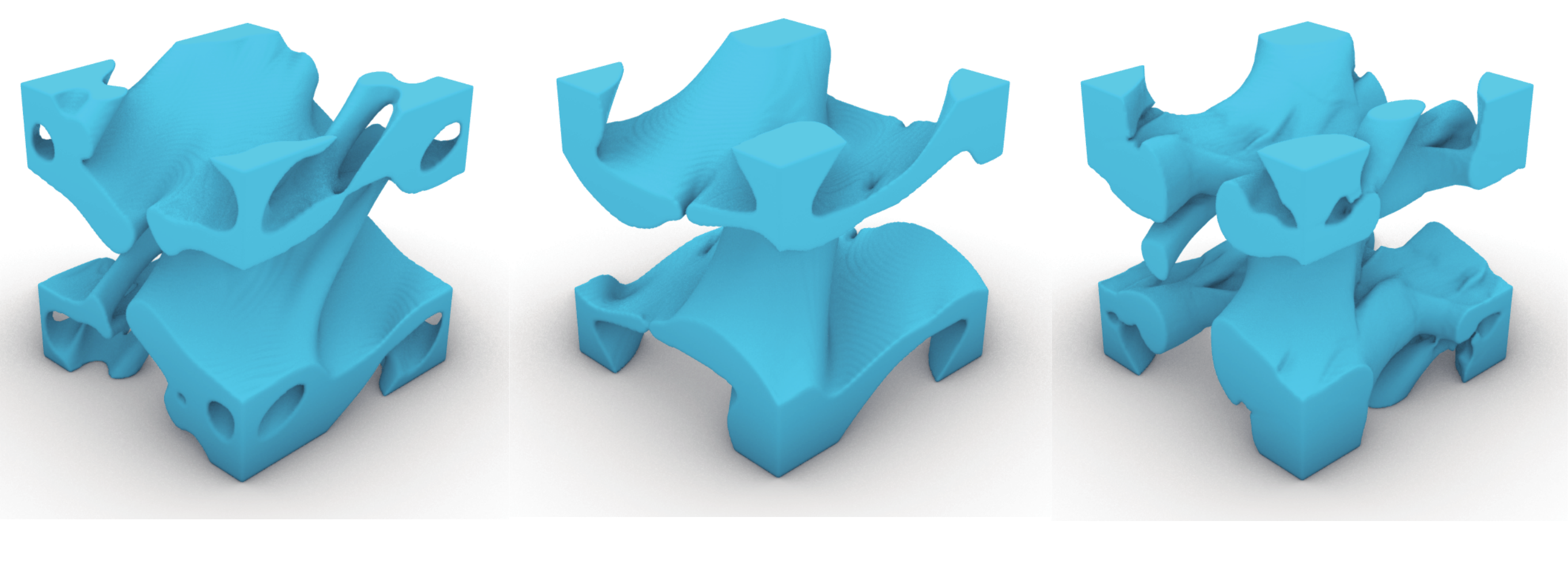}
		{
			\put(4.5,0.5){ Objective~\eqref{eq:mse}}
			\put(39,0.5){ Objective~\eqref{eq:fracmse}}
			\put(74,0.5){ Objective~\eqref{eq:l1}}
		}
	\end{overpic}
	\vspace{0mm}
	\caption{
		Comparisons of different objective functions. The target tensor is $[0.3, 0.2, 0.1, 0.05, 0.05, 0.05]$. The resulting thermal conductivity tensors from left to right are $[0.293, 0.198, 0.097, 0.043, 0.050, 0.046]$, $[0.286, 0.190, 0.095, 0.046, 0.048, 0.039]$, $[0.274, 0.198, 0.098, 0.050, 0.048, 0.048]$.
	}
	\label{fig:different objective}
\end{figure}
The corresponding code is:
\begin{lstlisting}
auto objective = (Hh(0, 0)-tt[0]).abs() + (Hh(1, 1)-tt[1]).abs() + (Hh(2, 2)-tt[2]).abs() + (Hh(0, 1)-tt[3]).abs() + (Hh(2, 1)-tt[4]).abs() + (Hh(0, 2)-tt[5]).abs();
\end{lstlisting}

Specifically, we set $\frac{\partial|a|}{\partial a} = 0$ when $a=0$. Fig.~\ref{fig:different objective} shows the optimization structures and their corresponding tensors under different objective functions.
From the results, there is no particularly noticeable difference in the performance between~\eqref{eq:mse} and~\eqref{eq:fracmse}. 
However, the results produced under the guidance of different objective functions exhibit variations. 
Users can customize the objective function for different applications. 


\paragraph{Different devices}

\begin{figure}[t]
	\centering
	\begin{overpic}[width=0.99\linewidth]{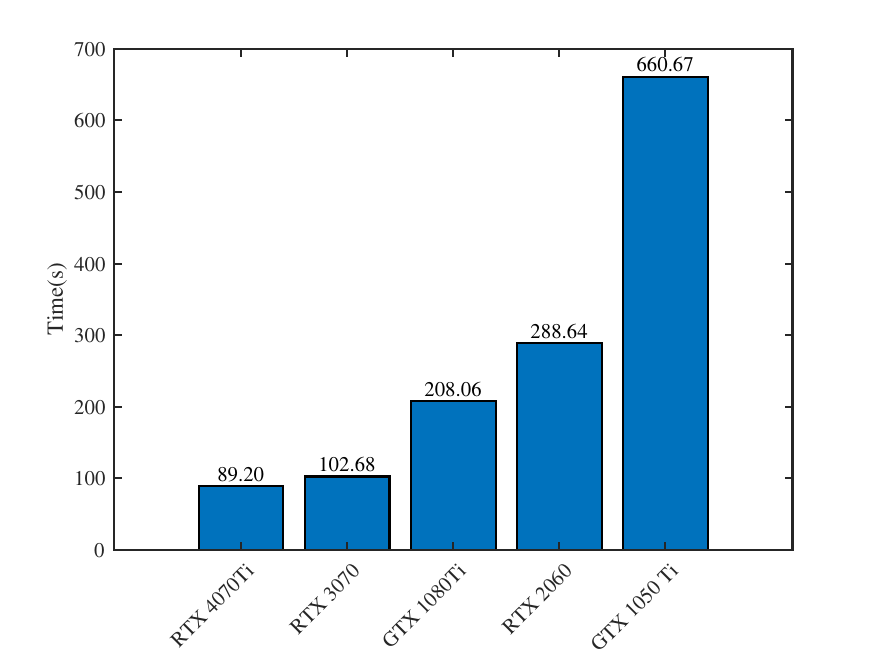}
		{
		}
	\end{overpic}
	\vspace{0mm}
	\caption{
		Five different devices. The RTX 3070 and RTX 2060 are tested on laptops, while the other devices are tested on desktops. The optimization model is~\eqref{md:model3} with a resolution of $128 \times 128 \times 128$.
	}
	\label{fig:time&machine}
\end{figure}

We test our framework on five devices, ranging from the 1050Ti to the 4070Ti, to ensure that it generates results with sufficient resolution on devices with general configurations. 
The results are presented in Fig.~\ref{fig:time&machine}. 
The chart demonstrates that the framework can generate a structure with the latest graphics cards within a reasonable waiting time. 
Moreover, with higher-performance graphics cards, it can generate a $128^3$ resolution structure in less than ninety seconds.

\section{Conclusion}\label{sec:conclusion}

We propose an open-source, user-friendly, large-scale thermal microstructure design framework using a single GPU. 
It offers various interfaces tailored to different users and has been tested on multiple devices to ensure successful operations. 
The fully open-source framework, along with a detailed installation video, can be downloaded from the following link~\url{https://github.com/quanyuchen2000/OPENTM}.
We offer customized optimization for specified thermal tensor matrices and propose an optimization model with adaptive volume constraints. 
The proposed method is robust and applicable to 2D and high-resolution 3D problems.
Moreover, the framework supports various optimization constraints, including isotropy, orthotropic anisotropy, and central symmetry. 
It can also be extended to various objective functions to meet application requirements. 

In future work, we will conduct in-depth research on more practical thermal problems using the thermal framework proposed in this paper. We will also explore the thermal behavior of microstructures to guide the material design of more advanced microstructures.

\paragraph{Acknowledgement}
This work is supported by the Youth Innovation Key Research Funds for the Central Universities, China (YD0010002010), the Open Project Program of the State Key Laboratory of CAD\&CG, Zhejiang University (Grant No. A2303).

\section*{Declarations}

\paragraph{Conflict of interest} The authors declare that they have no conflict of interest. 

\paragraph{Replication of results} Important details for replication of results have
been described in the manuscript.  Code for this paper is at ~\url{https://github.com/quanyuchen2000/OPENTM}.

\bibliographystyle{spbasic}    
\bibliography{ref}

\end{document}